\documentstyle[12pt]{article}
\newskip\humongous \humongous=0pt plus 1000pt minus 1000pt

\def\caja{\mathsurround=0pt}
\def\eqalign#1{\,\vcenter{\openup1\jot \caja
        \ialign{\strut \hfil$\displaystyle{##}$&$
        \displaystyle{{}##}$\hfil\crcr#1\crcr}}\,}
\newif\ifdtup

\renewcommand{\theequation}{\thesection.\arabic{equation}}
\newcommand{\beq}{\begin{equation}}
\newcommand{\eeq}[1]{\label{#1}\end{equation}}
\newcommand{\ber}{\begin{eqnarray}}
\newcommand{\eer}[1]{\label{#1}\end{eqnarray}}
\newcommand{\sect}[1]{\setcounter{equation}{0}\setcounter{footnote}{0}%
	\section{#1}}
\newcommand{\subsect}[1]{\setcounter{footnote}{0}\subsection{#1}}
\renewcommand{\a}{\alpha}
\renewcommand{\b}{\beta}

\renewcommand{\d}{\delta}

\newcommand{\g}{\gamma}

\newcommand{\e}{\epsilon}

\renewcommand{\l}{\lambda}

\newcommand{\m}{\mu}

\newcommand{\n}{\nu}

\newcommand{\s}{\sigma}

\def\ind{{(a)}}
\def\del{\partial}
\def\mod{\Phi}
\def\modb{\bar \Phi}
\def\mat{C}
\def\matb{\bar C}
\def\mpl{M_{\rm Pl}}
\def\beq{\begin{equation}}
\def\eeq{\end{equation}}
\def\Fc{{\cal F}}
\def\Fctree{{\cal F}^{(0)}}
\def\Fcol{{\cal F}^{(1)}}

\def\fone{h}
\def\ftwo{f}
\def\fonetree{h^{(0)}}
\def\ftwotree{f^{(0)}}
\def\foneol{h^{(1)}}
\def\ftwool{f^{(1)}}
\def\Hone{H^{(1)}}

\def\P{\hat{X}}
\def\Q{\hat{F}}

\def\Khat{{K_{\Phi}}}
\def\shift{\Xi}
\def\Ll{L}
\def\nws{N}
\def\hath{\sigma}
\def\hatV{V^{\rm inv}}
\def\hatS{S^{\rm inv}}

\begin{document}

\begin{titlepage}
\begin{center}

\hfill THU--95/5\\
\hfill UTTG--8-95\\
\hfill IASSNS--HEP--95/16\\
\hfill HUB--EP--95/2\\
\hfill hep-th/9504006\\

\vskip .6in

{\large \bf
Perturbative Couplings of Vector Multiplets in $N=2$
Heterotic String Vacua}\footnote{%
	Research supported in part by:
	the NSF, under grant PHY--90--09850 (V.~K.);
	the Robert A.~Welch Foundation (V.~K.);
	the Heisenberg Fellowship of the DFG (J.~L.);
	the NATO, under grant CRG~931380 (V.~K.$\&$J.~L.);
	the European Community Research Programme under contract
	SC1--CT92--0789 (B.d.W.\&J.L.).%
	}
\vskip .5in

{\bf Bernard de Wit$^1$, Vadim Kaplunovsky$^2$, Jan Louis$^{3,4}$
and Dieter L\"ust$^5$}
\footnote{email: bdewit@fys.ruu.nl,
	vadim@bolvan.ph.utexas.edu,
	luest@qft1.physik.hu-berlin.de,\\
	jlouis@lswes8.ls-wess.physik.uni-muenchen.de
	}
\\
\vskip 1.2cm
$^1${\em Institute for Theoretical Physics, Utrecht University \\
 Princetonplein 5, 3508 TA Utrecht, The Netherlands}

$^2${\em Theory Group, Dept.~of Physics, University of Texas\\
	Austin, TX 78712, USA}

$^3${\em Sektion Physik, Universit\"at M\"unchen\\
	Theresienstrasse 37, D-80333 M\"unchen, Germany}

$^4${\em Institute for Advanced Study,\\
 Olden Lane, Princeton, NJ 08540, USA}

$^5${\em Humboldt-Universit\"at zu Berlin,
Institut f\"ur Physik\\
D-10099 Berlin, Germany}\\

\end{center}

\vskip .3in

\begin{center} {\bf ABSTRACT } \end{center}
\vspace{-2mm}
We study the low-energy effective Lagrangian of $N=2$ heterotic
string vacua at the classical and quantum level.
The couplings of the vector multiplets are uniquely determined
at the tree level, while the loop corrections are severely constrained
by the exact discrete symmetries of the string vacuum.
We evaluate the general transformation law of the
perturbative prepotential and  determine its  form
for the toroidal compactifications
of six-dimensional $N=1$ supersymmetric vacua.

\vskip 1.5cm
April 1995
\end{titlepage}
\def\baselinestretch{1.2}
\baselineskip 16 pt
\setcounter{equation}{0}

\sect{Introduction and Summary}

Four-dimensional gauge theories invariant under $N=2$
supersymmetry have  revealed several interesting features
about (supersymmetric) quantum field theories, although they
themselves are not directly related to physical phenomena at or
below the electro-weak scale. In part these features are due to
the strong restrictions  imposed by $N=2$ supersymmetry on the
couplings of the classical Lagrangian and its
possible counterterms at the quantum level.
This fact leads to a number of non-renormalization theorems
which are usually much stronger than their $N=1$
counterparts \cite{NRT,NS}.
Recently, Seiberg and Witten \cite{SeibWit} were able to
solve rigidly $N=2$ supersymmetric Yang-Mills theory with a gauge
group $SU(2)$ using the analytic properties of
the $N=2$ couplings.
This led to exciting insight into non-perturbative
phenomena of quantum field theories, such as  confinement and
monopole condensation.

It is of  interest to consider the analysis of Seiberg and Witten and
its generalizations to larger gauge groups \cite{LKTY}
in the context of  $N=2$ string theories. Locally  $N=2$
supersymmetric gauge theories arise in four space-time dimensions
either as  type-II string vacua or as vacua of the heterotic
string. In all
type-II vacua the size of the gauge group is severely limited
\cite{DKV}, while in the heterotic string this constraint is much weaker
and large gauge groups typically appear at certain points in the
classical moduli space.
Therefore, the latter provides
the suitable framework to study non-trivial gauge dynamics \`a la
Seiberg and Witten. However, in  this paper we solely focus on the
the perturbative properties of  $N=2$  heterotic string vacua
as a first step in this direction.

In section~2 we briefly summarize known properties of the
four-dimensional $N=2$ heterotic strings.
{}From the world-sheet point of view, the heterotic vacua
do not have left-right symmetry:
their left-moving degrees of freedom are described in terms of a bosonic
conformal field theory (CFT) with central charge $\bar c=22$,
while the right-moving sector is  build out of
a free superconformal field theory (SCFT) with $c=3$ and an
interacting SCFT with $N=4$ world-sheet supersymmetry and  central
charge $c = 6$ \cite{BD}.
In space-time, the massless spectrum of such a vacuum always
contains the $N=2$ gravitational multiplet
consisting of the graviton, two gravitinos and the graviphoton
--- a spin-1 Abelian  gauge boson.
The other gauge bosons are members of
$N=2$ vector multiplets, which also contain two gauginos and a
complex scalar field, both in the adjoint representation of the gauge
group $G$.
The  matter fields which are charged under the gauge group
(usually in the fundamental representation of $G$)
reside in $N=2$ hypermultiplets.
Most string vacua also contain gauge-neutral moduli scalars,
which correspond to exact flat directions of the effective
potential. Their vacuum expectation values are not determined
at the perturbative level and are thus free parameters
of the string vacua. In $N=2$ theories such moduli can arise in
either vector or hypermultiplets.

The dilaton, the antisymmetric tensor and a
vector boson are contained in a special $N=2$ moduli
multiplet. In appendix A we display the linearized transformation
rules of this new $N=2$  multiplet,
called vector-tensor multiplet, as well as its free action, which
describes the appropriate degrees of freedom.
It is constructed out of an $N=1$ vector multiplet
and an $N=1$ linear multiplet
and its superalgebra necessarily has an off-shell central
charge. In four space-time dimensions an antisymmetric tensor is
always dual to a scalar field (the axion) and thus the
dilaton/axion can be put into an Abelian vector multiplet.
However, at present we can perform the duality transformation
only at the component level, without preserving the off-shell
supersymmetry.\footnote{Note that in type-II theories the dilaton and
the antisymmetric tensor reside in an $N=2$ tensor
multiplet, which is dual to an $N=2$ hypermultiplet
\cite{CecFerGir,linear}.}

Section~3 is devoted to locally $N=2$ supersymmetric field
theories at the classical and quantum level.
The  couplings of the vector multiplets
are encoded in a single holomorphic prepotential  $F$
that is homogeneous of degree two. The $\sigma$-model metric for
the scalar fields is the metric of a `special K\"ahler manifold'
in that its  K\"ahler potential is determined by
the holomorphic $F$ \cite{DWVP}; the gauge couplings follow from
the second  derivatives of $F$.
The global symmetries of the theory, continuous or discrete,
have to respect the special  properties of the
vector couplings
and therefore have to constitute a subgroup of  $Sp(2n+2)$,
where $n$ is the number of vector multiplets.

In section~3.1 we briefly review the couplings of the vector
multiplets and their transformation properties; this  will  be
important for discussing the
discrete quantum symmetries of the heterotic string in
section~4.
We draw attention to the fact that in certain
parametrizations a prepotential does not exist, a phenomenon
discussed recently in \cite{Ceresole}; the relevance of such
parametrizations is discussed later.
In section~3.2 we study quantum properties of $N=2$ theories.
Just as in $N=1$ supersymmetric quantum field theories, it
is essential  to distinguish the Wilsonian couplings
of the effective theory from the physical, momentum-dependent
effective couplings \cite{SV,KLa}.
The former share all the analytic properties of their
classical counterparts; in particular, they
are determined from the loop corrections to the holomorphic
prepotential $F$.
But in theories with massless charged fields,
the momentum-dependent effective gauge couplings
are different from the Wilsonian couplings
and do not share their analytic properties.
The difference between the two kinds of couplings can be computed entirely
within the low-energy effective field theory \cite{HOLAN,KLa,KLb}; for
$N=2$ this computation  is outlined in section~3.2.

In any heterotic string vacuum, the dilaton enjoys
very special properties.
To all orders in perturbation theory it
has a continuous Peccei-Quinn symmetry,
and at the tree level, it is completely orthogonal to the other
moduli scalars.
According to a theorem by Ferrara and Van Proeyen \cite{FVP},
these two requirements uniquely determine the {\em tree-level}
prepotential $F$ for all heterotic $N=2$ vacua.
The properties of this prepotential, in
particular its target-space symmetries, are discussed in
section~4.1.
(Most of these tree-level results were independently
obtained in ref.~\cite{Ceresole}).

The loop corrections to the prepotential are severely constrained
by the Peccei-Quinn symmetry, which prohibits any higher-loop
corrections, and by the exact target-space duality symmetries
of the string vacua. In section~4.2 we generalize the formalism
of section~3.1 to include quantum corrections
and determine the general transformation law
of the loop-corrected $F$.

Finally, in section~4.3 we apply this formalism to the concrete case
of $N=2$ vacua that arise as toroidal compactifications
of six-dimensional $N=1$ theories.
This class of vacua always features the two toroidal moduli $T$
and $U$ and an exact invariance
under the modular group  $[SL(2,{\bf Z})]^2$.
For this case the modular forms are known and the
transformation laws derived in  section~4.2 completely determine the
loop-corrected prepotential
(up to a quadratic polynomial that may add field-independent constants to
the gauge couplings).
We also discuss what happens when a toroidal compactification
is deformed by a Wilson-line modulus.

\sect{$N=2$ Vacua of the  Heterotic String}

In this section we recall some of the known feature of four-dimensional
$N=2$  heterotic string vacua which are  needed for our later
discussions. Further details can be found in the literature,
for example, in refs.~\cite{BD,LLT}.

One way to characterize different string vacua is via their
underlying two-dimensional (super)conformal field theories
(SCFTs) on the world sheet.
As we have already mentioned, the SCFTs of the heterotic vacua
are not left--right symmetric and their left-moving and right-moving
world-sheet degrees of freedom are quite distinct from each other.
The right-moving  side has a local world-sheet
supersymmetry and consists of  four free bosons
$X^\m(z)$, which, along  with  four free fermions $\psi^\m(z)$,
generate the four-dimensional space-time. Together with the
superconformal ghosts $b(z)$, $c(z)$, $\b(z)$ and $\g(z)$, they
contribute $-9$ units to the central charge which has to
 be balanced by the central charge of an appropriate ``internal"
SCFT. This internal SCFT is further constrained by the desired
space-time properties
of the string vacuum,  in particular by the amount of
space-time supersymmetry.

$N$-extended space-time supersymmetry implies the existence
of $N$ supercharges $Q_\alpha^i$ ($i=1,\dots , N$) obeying the
supersymmetry algebra in four-dimensional Minkowski space:
\begin{equation}
\eqalign{
\lbrace Q_\alpha^i,\bar Q_{\dot\beta j}\rbrace & =2i\delta^i_j
\gamma^\mu_{\alpha\dot\beta}\,P_\mu \,, \cr
\lbrace Q_\alpha^i,Q_\beta^j\rbrace & =2C_{\alpha\beta}\,Z^{ij}\, ,\cr}
\end{equation}
where $Z^{ij}$ denotes the  central charges.
 $\alpha$ and $\dot\beta$ are two-component  spinor
indices associated with chiral spinors in a four-dimensional
space-time, and $C_{\a\b}$ is the charge-conjugation matrix. In
the heterotic string theory, all the supercharges `live' on the
right-moving (supersymmetric)  side of the SCFT and thus
are defined in terms of holomorphic local two-dimensional
operators given by the holomorphic parts of
the gravitino vertex operators $V_\alpha^i$ and $V_{\dot\alpha i}$
at zero momentum:
\begin{equation}
Q_\alpha^i=\oint {{\rm d}z\over
2\pi i}V_\alpha^i(z)\,,\qquad \bar Q_{\dot\alpha i}=\oint
{{\rm d}z\over 2\pi i}\bar V_{\dot\alpha i}(z)\,,
\end{equation}
where  $V_\alpha^i(z)$ and $\bar V_{\dot\alpha i}(z)$
(in the $-1/2$ ghost picture) are given by
\begin{equation}
V_\alpha^i(z)=S_\alpha\Sigma^ie^{-{\phi\over 2}}(z)\,,\qquad
\bar V_{\dot\alpha i}(z)=S_{\dot\alpha}\Sigma_ie^{-{\phi\over
2}}(z)\,.
\end{equation}
Here $S_\alpha(z)$, $S_{\dot\alpha}(z)$ are four-dimensional spin
fields and $\phi(z)$ originates from the bosonization of the
superconformal ghosts fields $\b$ and $\g$.
$\Sigma^i(z)$ and $\Sigma_i(z)$ are conformal fields
of the $c=9$  internal SCFT and have  conformal dimension $3/8$.
For heterotic vacua the $N=2$ space-time supersymmetry charges reside
entirely in the right-moving sector. This is in contrast with the
type-II string, where one has one right-moving and one
left-moving supercharge operator.

For a  heterotic string vacuum  with  $N=2$ space-time supersymmetry
the algebra (2.1) implies that the internal right-moving  $c=9$
SCFT splits into a $c=3$ SCFT with $\nws=2$ world-sheet supersymmetry
and a $c=6$ piece with $\nws=4$ supersymmetry on the world-sheet
\cite{BD,LLT}. The $c=3$ system can be
realized by a free complex $\nws=1$ superfield
whose bosonic components we denote by $\partial X^\pm(z)$, while
the complex fermions are  bosonized according to
$\psi^\pm(z)=\exp(\pm iH(z))$ with
$H(z)$ a  free real scalar field.
On the other hand, the $\nws=4, c=6$ SCFT is not  free,
 although it necessarily contains a free boson
$J^3(z)={i\over \sqrt2}\partial H'(z)$ corresponding to the
Cartan generator of the level-1 $SU(2)$ Ka\v{c}-Moody algebra of the
$\nws=4$  theory. In terms of $H$ and $H'$
the internal part  of the gravitino  vertex
operators can be represented by
($\Sigma_j(z)=(\Sigma^j)^\dagger$):
\begin{equation}
\Sigma^{j}(z)=\exp\Big({i\over 2}H(z)\Big)\;\exp\Big((-)^{j+1}
{i\over\sqrt2}{H'}(z)\Big)\,.
\end{equation}
Evaluating the operator product of $\Sigma^i(z)$ with $\Sigma^j(w)$
one finds  the central-charge operator of the $N=2$  algebra to be
\begin{equation}
Z^\pm=\oint{{\rm d}z\over 2\pi i}\;\psi^\pm (z)\,e^{-\phi(z)}.
\end{equation}
In the zero-ghost picture $Z^\pm$ is
proportional to  $\partial X^\pm (z)$ and hence
the central charge of a massive state is determined by the
momentum eigenvalues ($p^+,p^-$) in the $c=3$ SCFT.
This implies that vertex operators for  massive states with
non-vanishing central charges
contain factors of the form $\exp\big(ip^+X^-(z)+ip^-X^+(z)\big)$.

Generally, there is no world-sheet supersymmetry on the
left-moving side of the heterotic string, which
is based on an ordinary  bosonic CFT with
central charge $\bar c = 26$.
This CFT is comprised of a $\bar c
=4$ sector containing four free bosons $X^{\mu}(\bar z)$, which generate
the four-dimensional space-time, and of an arbitrary  $\bar c =22$ sector.

The massless spectrum of a heterotic $N=2$ vacuum
always comprises the graviton
($G_{\mu\nu}$), the antisymmetric tensor
($B_{\mu\nu}$) and the dilaton ($D$), created by vertex
operators of the form $\bar\partial X_\mu(\bar z)\,
\del X_\nu(z)$ (at zero momentum),  and two gravitini and two
dilatini, created by vertex operators $\bar\partial X_\mu(\bar
z)\,V^i_\alpha (z)$. In addition, there are
always two Abelian gauge bosons $A_\mu^\pm$ with vertex operators
$\bar\partial X_\mu(\bar z)\,\partial X^\pm(z)$, which  generate
the  gauge group $[U(1)_R]^2$ (the suffix $R$
indicates that these groups originate from the dimension-one
operators $\partial X^\pm(z)$ of the right-moving
sector). One linear combination is  the graviphoton, which is the
spin-1 gauge boson of the $N=2$ supergravity multiplet that also
contains the graviton and two gravitini.\footnote{%
    Note that the  $\partial X^\pm(z)$ factors
    of the vertex  operators of the Abelian gauge bosons coincide
    with  the central-charge operator  and thus the
    charges of $[U(1)_R]^2$ are identical to the central charges
    $p^+,p^-$.}
The dilaton together
with the antisymmetric tensor $B_{\mu\nu}$, the two dilatini
and the remaining $U(1)$ vector are naturally described by an
$N=2$ {\it vector-tensor}  multiplet.
As far as we know,
this type of an $N=2$ supermultiplet has not been discussed in
the previous literature.
In appendix A we establish the linearized transformation properties
of this new  vector-tensor  multiplet and display its free action.
In a dual description, where the antisymmetric tensor is replaced by
a pseudo-scalar (axion)  $a$, the  degrees of freedom
form a  $N=2$ vector supermultiplet where
the dilaton  and the axion
combine into  a complex scalar $S =e^D+ia$.\footnote{%
    At present we can perform the
    duality transformation  only  at the component level.}
Note that the `heterotic'
vector-tensor multiplet is different from the standard $N=2$ tensor
multiplet \cite{linear}, which describes the dilaton multiplet in
type-II $N=2$ theories.
The `type-II' tensor multiplet contains
no vector boson but has instead two additional
Ramond-Ramond scalars and is thus dual to an
$N=2$ hypermultiplet \cite{CecFerGir}.

Apart from the two Abelian gauge bosons we just discussed,
the massless spectrum of  heterotic string vacua   contain
further gauge bosons  $A_\mu^a$, which are always members of
$N=2$ vector multiplets. Their  superpartners are two gaugini
$\lambda^a_{i\, \alpha}$ and a complex scalar $\mat^a$ and
the vertex operators for a generic vector  multiplet
are given by
\beq
\Big(A_\mu^a,\lambda^a_{i\, \alpha},\mat^a \Big) \sim
\Big(J^a(\bar z)\, \del X_\mu (z),
J^a(\bar z)\, V_{i\, \alpha} (z),
J^a(\bar z)\, \del X^\pm (z)\Big)\,,
\eeq
where $J^a(\bar z)$ are dimension $(1,0)$ operators that together
comprise a left-moving Ka\v{c}-Moody current  algebra.
Their zero modes generate a non-Abelian gauge group $G$;
the currents $J^a$ and hence the entire
corresponding vector multiplets transform  in the adjoint
representation of this group.
The maximal rank of $G$ is bounded by the central
charge $\bar c$ of the Ka\v{c}-Moody algebra and therefore cannot
exceed $22$.
Furthermore,  $G$
is not necessarily a simple group,
but rather contains several  simple and/or Abelian factors.
For example, the compactification of the ten-dimensional
heterotic string on a six-dimensional manifold $T^2\times K_3$
leads to the gauge group
$E_7\times E_8\times U(1)_L^2 \times  U(1)_R^2$\footnote{%
    In the orbifold limit of $K_3$ there is an additional $SU(2)$ factor.
    Furthermore, for special values of the orbifold's radii,
    there are up to four additional $SU(2)$ factor, so the maximal
    gauge group of a $T^2\times K_3$ compactification is
    $E_7\times E_8\times SU(2)^5\times U(1)_L^2 \times  U(1)_R^2$.}
and the world-sheet supersymmetries are maximally
extended to $(2,2)\oplus (4,4)$.
Altogether, this class of vacuum families has ${\rm dim}(E_7)+
{\rm dim}(E_8) + 2+2 =385$  gauge fields and the corresponding
low-energy effective $N=2$ supergravity has 384 vector supermultiplets
(including the Abelian vector multiplet for the dilaton but excluding
the graviphoton).

The scalar fields in the Cartan subalgebras of
non-Abelian factors $G_\ind\subset G$ as well as
the scalars of any Abelian factor in  $G$
correspond to
flat directions of the $N=2$ scalar potential.
Their vertex operators  are truly marginal operators
of the SCFT and
the corresponding space-time vacuum expectation values
are free parameters which
continuously connect a family of string vacua.
However, the
low-energy description of the  vacuum family depends
on the size of these vacuum expectation values.
If they vanish,
the entire non-Abelian gauge group
is intact and the associated massless gauge bosons
appear in the low-energy effective Lagrangian.
On the other hand, non-vanishing vacuum expectation values of flat
directions would spontaneously break the gauge symmetry down
to some subgroup of $G$ and only the gauge fields in the adjoint
representation of this subgroup would remain massless.
When the vacuum expectation values of the flat directions
are large ($O(\mpl)$), the massive gauge
bosons and their superpartners are superheavy and hence should be
integrated out of the low-energy effective theory,
which would then contain only the left-over light
degrees of freedom.
Consequently,
the flat directions responsible for the gauge symmetry breaking
now reside in Abelian vector multiplets of the low-energy theory;
they are a subset of the moduli fields and
in this article we combine them with the dilaton field $S$ and
denote them collectively by  $\mod^\a$.
If all scalars in the Cartan
subalgebra of $G$ have  non-zero vacuum expectation values,
$G$ is broken down to its maximal Abelian subgroup,
which can be at most
$U(1)^{22}$;  in this  case the  dimension of the moduli space
spanned by the vacuum expectation values of $\mod^\a$ would be
$22+1$ (the additional modulus corresponds to $S$).

The  matter fields which are charged under the gauge group
are generated by primary (chiral) operators in the right-moving
$N=4,\ c=6$ SCFT and they are members of  $N=2$ hypermultiplets.
Frequently, this right-moving SCFT also has truly marginal
directions that come in $N=2$ hypermultiplets; they do not mix with
the moduli belonging to vector multiplets and therefore
span a separate, orthogonal component of the total moduli space.
For example,   compactifications on the   $K_3$ surface  have
80 additional moduli scalars, which are gauge singlets
and reside in 20  $N=2$ hypermultiplets.
Their vacuum expectation values
span the  moduli space of $K_3$  \cite{Seiberg},
\begin{equation}
{\cal M}_{K_3}={SO(20,4)\over SO(20)\times SO(4)}\,.
\label{modk3}
\end{equation}
However, for generic string vacua the moduli space of the
hypermultiplets is  unknown.

In later sections of  this paper we focus on the particular subclass of
four-dimensional  $N=2$  heterotic vacua, namely
compactifications of  six-dimensional $N=1$ heterotic vacua on a
two-torus~$T^2$.
The right-moving  coordinates of the torus are  given by the  operators
$\partial X^\pm(z)$ discussed   previously, but now
there also exist two  free complex left-moving operators
$\bar{\partial} X^\pm(\bar z)$, which can be  used to build
vertex operators for the two complex  moduli   of the torus
$\bar{\partial} X^\pm(\bar z)\,\del X^\pm(z)$.
 The moduli of $T^2$ are commonly
denoted by $T=2(\sqrt G+iB)$ and $U=(\sqrt G-iG_{12})/G_{11}$,
where $G_{ij}$ is the metric of $T^2$, $\sqrt G$ its determinant and $B$
the constant antisymmetric-tensor background; $U$ describes
the deformations of the complex structure while  $T$
parameterizes the deformations of the area and the antisymmetric tensor,
respectively.
The moduli space spanned by  $T$ and $U$ is determined by
the Narain lattice of $T^2$ \cite{NAR}:
\begin{equation}
{\cal M}_{T,U}=
\biggl({SO(2,2)\over SO(2)\times SO(2)}\biggr)_{T,U}
\simeq
\biggl({SU(1,
1)\over SU(1)}\biggr)_T \otimes \biggl({SU(1,1)\over
SU(1)}\biggr)_U.
\end{equation}
All physical properties of the two-torus compactifications are invariant
under the group $SO(2,2,{\bf Z})$ of discrete duality
transformations \cite{DUAL}, which comprise the $T\leftrightarrow U$
exchange and the
$PSL(2,{\bf Z})_{T}\times PSL(2,{\bf Z})_U$ dualities, which
acts on $T$ and $U$ as
\begin{equation} T\rightarrow {aT-ib\over icT+d}\,,\qquad
U\rightarrow{a'U-ib'\over ic'U+d'}\,,\label{tutransf}
\end{equation}
where the parameters $a,\ldots,d'$ are integers and  constrained by
$ad-bc=a'd'-b'c'=1$.

$T$ and $U$ are the spin-zero
components of two additional $U(1)$ $N=2$ vector supermultiplets.
The necessary
enlargement of the Abelian gauge symmetry is furnished by vertex
operators of the form $\bar\partial
X^\pm(\bar z)\,\del X_\mu(z)$ which generate
the  gauge group $[U(1)_L]^2$.
This results in a combined  gauge
symmetry of  $[U(1)_{+}]^2\times [U(1)_{-}]^2$, where the subscript
$\pm$ indicates the combinations $L\pm R$. The $[U(1)_+]^2$ originates
from the internal graviton mode of the six-dimensional theory
compactified on $T^2$ while   $[U(1)_-]^2$ originates from
the compactification of the six-dimensional antisymmetric tensor
field. In general, the four Abelian gauge bosons transform into
each other under target-space duality transformations.\footnote{For
example, the transformation $T\rightarrow 1/T$ rotates the left-
and right-moving torus coordinates since it involves an exchange
of momentum and winding numbers (see \cite{HIGGS} for details).
This can be seen as a transformation on  the world-sheet
electric and magnetic charges. In particular, for
the simple case of $U=1$, ${\rm Im}\,T=0$, the transformation
$T\rightarrow 1/T$  just acts as $\partial X^\pm (z)\rightarrow
-\partial X^\pm (z)$ with
$\bar\partial X^\pm (\bar z)$  invariant and hence the groups
$[U(1)_+]^2$ and  $[U(1)_-]^2$ are simply interchanged.}

At special points in the $(T,U)$ moduli space, additional
vector fields become massless and the $U(1)^2_L$ becomes
enlarged to a non-Abelian gauge symmetry.
In particular, along the critical $T=U$ line, there are two additional
massless gauge fields and the $U(1)^2_L$ becomes  $[SU(2)\times U(1)]_L$.
The scalar superpartners of the three gauge bosons of the $SU(2)_L$
include $a=T-U$, which acts as the Higgs field breaking the $SU(2)_L$
when one moves away from the $T=U$ line~\cite{HIGGS}.
Similar critical lines exist for $T\equiv U\!\!\pmod{SL(2,{\bf Z})}$,
$i.\,e.$, $T=(aU-ib)/(icU+d)$ for some integer $a,b,c,d$ with $ad-bc=1$.
When two such lines intersect, each line brings with it a pair of massless
gauge fields and the gauge symmetry becomes enhanced even further;
the enhanced group may be determined by simply counting
the intersecting critical lines~\cite{CLM1}.
For example, the point $T=U=1$ lies at the intersection of two
critical lines, namely $T=U$ and $T=1/U$, and hence has four extra gauge
bosons.
The corresponding gauge symmetry is $SU(2)^2_L$ and the two Higgs
fields
in the Cartan subalgebra of this symmetry can be identified as
$a_1=T-U$ and $a_2=T-(1/U)$\footnote{Alternatively, the Higgs fields can be
defined as $a_1'={T-1\over T+1}$, $a_2'={U-1\over U+1}$;
$({T-1\over T+1})^2$ and $({U-1\over U+1})^2$ correspond
to the uniformizing variables of modular functions around the critical points
$T=1$ and $U=1$. For the case an enhanced $SU(3)_L$ gaue group the analogous
definitions are $a_1'={T-\rho\over T+\bar\rho}$ and $a_2'={U-\rho\over
U+\bar \rho}$.}.
Similar two-line intersections happen whenever
$T\equiv U\equiv 1\!\!\pmod{SL(2,{\bf Z})}$ and the gauge group
is enhanced to an $SU(2)^2_L$ at all such points.
On the other hand, three critical lines $T=U$, $T=1/(U-i)$ and
$T=(iU+1)/U$ intersect at the critical point
$T=U=\rho=e^{2\pi i/12}$, where one therefore has six massless gauge
bosons in addition to the $U(1)^2_L$;
this enhances the gauge symmetry all the way to an $SU(3)_L$.
Two Higgs scalars in its Cartan subalgebra of this symmetry can be
identified as {\it e.~g.}, $a_1=T-U$ and $a_2=((iT+1)/T)-(1/(U-i))$.
Again, similar triple intersections occur at
$T\equiv U\equiv \rho\!\!\pmod{SL(2,{\bf Z})}$ and the gauge group
is enhanced to an $SU(3)_L$ at all such points.
The above is the complete list of all the critical lines and points
of the $(T,U)$ moduli space; at all the other point, the $(T,U)$
system has only the $U(1)^2_L$ gauge symmetry.
In particular, there is no enlargement of the gauge symmetry when
$T\equiv1$ but $U\not\equiv1$ or $T\equiv\rho$ but $U\not\equiv\rho$;
this fact will be important for our analysis in section 4.3.

At the critical points, complete $N=2$ supermultiplets become
massless and the vertex operators of their  bosonic components
$(A_\mu, \mat^\pm)$ have the  form
\begin{equation}
\eqalign{
&A_\mu\sim e^{i(\bar p^+(T,U)\bar X^-(\bar z)+\bar p^-(T,U)\bar
X^+(\bar z))} \,
e^{i(p^+(T,U)X^-(z)+p^-(T,U)X^+(z))}\,\del X_\mu(z),\cr
&\mat^\pm\sim e^{i(\bar p^+(T,U)\bar X^-(\bar z)+\bar p^-
(T,U)\bar X^+(\bar z))}\,
e^{i(p^+(T,U)X^-(z)+p^-(T,U)X^+(z))}\,\del X^\pm(z),\cr}
\end{equation}
where the Narain lattice vectors satisfy
\begin{equation}
\bar p^+(T,U)\,\bar p^-(T,U)= p^+(T,U)\,p^-(T,U)+2\ .
\end{equation}
The masses of such  states are given by $m^2(T,U)={1\over
2}p^+(T,U)\,p^-(T,U)$, which indeed vanish precisely
at the critical points.
Away from the critical points, the multiplet
$(A_\mu, \mat^\pm)$ has
non-vanishing right-moving lattice momentum
vectors $(p^+,p^-)$, which implies that these massive states have
non-vanishing central charges. Therefore they build small
representations of the $N=2$ supersymmetry algebra.

Toroidal compactifications can be continuously deformed by turning
on non-trivial Wilson lines in the gauge group for the two periods
of the two-torus.
Such deformations give rise to additional moduli belonging to
Abelian vector supermultiplets; we denote such Wilson-line moduli
by $\phi^i$ with $i=1,\ldots,P$ ($P\le20$).
The  combined  moduli space spanned by $T$, $U$ and $\phi^i$
can be directly derived  from  the Narain lattice
of the heterotic string compactification on $T^2$ \cite{NAR,CLM2}.
One finds  the   symmetric K\"ahler space
(see also section 4):
\begin{equation}
{\cal M}_{T,U,\phi^i} =\biggl({SO(2,P+2)\over
SO(2)\times SO(P+2)}\biggr)\Big/ SO(2,2+P,{\bf Z}) .
\end{equation}
(see also section 4).
Together with the dilaton field $S$, which parameterizes
the coset space $SU(1,1)/U(1)$, we are thus dealing
with an $(3+P)$--dimensional space, spanned by the complex moduli
$\Phi^\alpha$.
At a {\it generic} point in this moduli space
the gauge group is
$U(1)_{L+R}^2\times U(1)_{L-R}^2\times U(1)^{P}$;
it is enlarged to a non-Abelian gauge group at special points
(or rather subspaces) of the moduli space.

Target-space duality transformations now act simultaneously on all moduli
fields $T$, $U$ and the additional moduli $\phi^i, i=1,\ldots,P$.
Specifically, the target-space duality transformations are contained in
the discrete group $SO(2,2+n,{\bf Z})$, which possesses
$PSL(2,{\bf Z})_T\times PSL(2,{\bf Z})_U\subset
SO(2,2,{\bf Z})$ as a subgroup.
For example, $PSL(2,{\bf Z})_T$ acts on $T$
in the standard way (see eq.(\ref{tutransf}));
the $\phi^i$ transform with modular weight
$-1$ under this transformation, {\it i.e.},
\begin{equation}\phi^i\rightarrow {\phi^i\over icT+d}.
\label{phitrans}
\end{equation}
However $U$ transforms also non-trivially under this transformation as
\cite{ANTO,CLM2}:
\begin{equation}
U\rightarrow U-{ic\over icT+d}{\phi^i\phi^i}.
\label{shift}
\end{equation}
Thus, in the presence of the $\phi^i$, $T$ and $U$ get mixed under
duality transformations, which is a reflection of the
non-factorizable structure of the moduli space.

\vskip 0pt plus 1in \penalty -1000
\sect{Effective Quantum Field Theories\hfil\hbox{}\penalty-9000\
    with Local $N=2$ Supersymmetry}

In this section we
summarize generic properties of the effective $N=2$
supergravity action with particular emphasis on the couplings
of the vector multiplets.
The section is divided into two parts;
in  3.1 we  summarize  and further
develop a number of useful results of
special geometry and in  3.2 we discuss
quantum effects in effective $N=2$ supersymmetric theories.

\subsect{Summary of special geometry}

\noindent
In $N=2$ supersymmetric Yang-Mills theory the action is encoded
in a holomorphic prepotential $F(X)$, where $X^A$ ($A=1,\ldots,n$)
denote the vector superfields and also the complex
scalar components of such superfields.
The function $F(X)$ is usually assumed to be
invariant under the gauge group, although this requirement is not
always necessary \cite{DWLVP,DWHR}.
Two different functions
$F(X)$ may correspond to equivalent equations of
motion; generically the
equivalence involves symplectic reparametrizations combined with
duality transformations, which we will turn to shortly.
The local $N=2$ supersymmetry requires an additional vector
superfield $X^0$ in order to accomodate the graviphoton, but the
scalar and the spinor components of this superfield do not
lead to additional physical particles.
Therefore, in the local case $F(X)$ is a holomorphic function of
$n+1$ complex variables $X^I$ ($I=0,1,\ldots,n$), but it
must be a {\em homogeneous} function of degree two  \cite{DWVP}.
According to the superconformal multiplet calculus, the physical scalar
fields of this system
parameterize an $n$-dimensional complex hypersurface,
defined by the condition that the imaginary part of $X^I\,\bar
F_I(\bar X)$ must be a constant linearly related to Planck's
constant, while the overall phase of the $X^I$ is irrelevant in
view of a local chiral invariance.
The embedding of this hypersurface can be described in terms of
$n$ complex coordinates $z^A$ by letting $X^I$ be proportional to
some holomorphic sections $X^I(z)$ of the projective space.
The resulting geometry for the space of physical scalar fields
belonging to vector multiplets of an $N=2$
supergravity is a {\em special} K\"ahler geometry \cite{DWVP,special},
with a K\"ahler metric $g_{AB}=\del_A\del_{\bar B}K(z,\bar z)$
following from a K\"ahler potential of the special
form\footnote{%
    Here and henceforth we use the standard convention where
    $F_{IJ\cdots}$ denote multiple derivatives with respect to $X$ of
    the holomorphic prepotential.
    }
\begin{equation}
K(z,\bar z)=
-\log\Big(i \bar X^I(\bar z)\,F_I(X(z)) -i X^I(z)\,
\bar F_I(\bar X(\bar z))\Big) .
\label{KP}
\end{equation}
The curvature tensor associated with such a special K\"ahler space
satisfies the characteristic relation \cite{BEC}
\begin{equation}
R^A{}_{\!\!BC}{}^{\!D} = 2\d^A_{(B} \d^D_{C)}  - e^{2K}
{\cal W}_{BCE}\, \bar {\cal W}{}^{EAD}\, ,
\end{equation}
where
\begin{equation}
{\cal W}_{ABC} =  F_{IJK}\big(X(z)\big) \;{\partial
  X^I(z)\over \partial z^A}  {\partial X^J(z)\over \partial z^B}
{\partial X^K(z)\over \partial z^C} \,.
\end{equation}
Up to an irrelevant phase, the proportionality factor between
the $X^I$ and the holomorphic sections $X^I(z)$ is equal to
$\exp\big({1\over 2} K(z,\bar z)\big)$.
A convenient choice of inhomogeneous coordinates $z^A$
are the {\em special} coordinates, defined by
\begin{equation}
z^A =X^A/X^0,\qquad A=1,\ldots ,n,
\end{equation}
or, equivalently,
\begin{equation}
X^0(z)=1\,,\qquad X^A(z) = z^A\,.
\end{equation}
In this parameterization the K\"ahler potential can be written as
\cite{SU}
\begin{equation}
K(z,\bar z) = -\log\Big(2({\cal F}+ \bar{\cal F})-
           (z^A-\bar z^A)({\cal F}_A-\bar{\cal F}_A)\Big)\,,
\label{Kspecial}
\end{equation}
where ${\cal F}(z)=i(X^0)^{-2}F(X)$.

The Lagrangian terms containing the kinetic energies of
the gauge fields are
\begin{equation}
{\cal L}^{\rm gauge}\ =\ -{\textstyle{i\over 8}}
\left( {\cal N}_{IJ}\,F_{\mu\nu}^{+I}F^{+\mu\nu J}\
-\ \bar{\cal N}_{IJ}\,F_{\mu\nu}^{-I} F^{-\mu\nu J} \right)\,,
\end{equation}
where $F^{\pm I}_{\m\n}$ denote the selfdual and anti-selfdual
field-strength components and
\begin{equation}
{\cal N}_{IJ}=\bar
F_{IJ}+2i {{\rm Im}(F_{IK})\,{\rm Im}(F_{JL})\,X^KX^L\over {\rm
Im}(F_{KL})\,X^KX^L} \,.
\label{Ndef}
\end{equation}
Hence $\cal N$ is the field-dependent tensor that comprises the
inverse gauge couplings
$g^{-2}_{IJ}= {i\over 4}({\cal N}_{IJ}-\bar {\cal N}_{IJ})$
and the generalized $\theta$ angles
$\theta_{IJ}= 2\pi^2({\cal N}_{IJ}+\bar {\cal N}_{IJ})$.
Note the important identity $F_I= {\cal N}_{IJ}\,X^J$.

As we already mentioned, different functions $F(X)$ can lead to
equivalent equations of motion. Such equivalence often involves
the electric-magnetic
duality of the field strengths rather than local transformations
of the vector potentials $A^I_\mu$.
For the non-Abelian case such a duality does not make
sense (because the field equations depend explicitly on the
vector potentials),
but it is perfectly legitimate in the context of Abelian gauge fields
when all the {\em fundamental} fields of the theory are neutral.%
\footnote{%
    A local fundamental field can be electrically charged but it cannot
    carry a magnetic charge.
    On the other hand, an extended object like a soliton can have
    both electric and magnetic charges.
    Therefore, when all the fundamental fields are neutral, one is free
    to choose any integral basis for the electric and magnetic charges,
    but a charged local field (in particular, a non-Abelian gauge field)
    restricts this choice since its charge must be electric rather
    than magnetic.
    }
With this proviso in mind, let us introduce the
duality transformations.
Following ref.~\cite{DWVP} and appendix C of ref.~\cite{DWVVP})%
\footnote{%
    As compared to the definitions in \cite{DWVP,BEC,DWVVP},
    our notation is as follows:
    $$
    \displaylines{
    [K(z,\bar z)]^{\rm here}= -K(z,\bar z) - \log 2\,,\qquad
    [{\cal W}_{ABC}]^{\rm here}= -2i Q_{ABC}\,. \cr
    [F(X)]^{\rm here} = -{\textstyle{i\over 2}} F(X)\,,\qquad
    [G^{+}_{\m\n I}]^{\rm here} = -iG^{+}_{\m\n I}\,.\qquad
    [{\cal N}_{IJ}(X,\bar X)]^{\rm here}
	= 2i {\cal N}_{IJ}(X,\bar X) \,, \cr }
    $$
    Note that the change in the K\"ahler potential induces a change
    of sign in the K\"ahler metric.
    },
we define the tensors $G^{\pm}_{\mu\nu I}$ as
\begin{equation}
G^+_{\mu\nu I}={\cal N}_{IJ}F^{+J}_{\mu\nu}\,,\quad G^-_{\mu\nu
I}=\bar{\cal N}_{IJ}F^{-J}_{\mu\nu}\,. \label{defG}
\end{equation}
Then the set of Bianchi identities and equations of motion
for the Abelian gauge fields can be
written as
\begin{equation}
\partial^\mu \big(F^{+I}_{\m\n} -F^{-I}_{\m\n}\big)
=0\,,\qquad
\partial^\mu \big(G_{\mu\nu I}^+ -G^-_{\m\n I}\big) =0\,,
\label{Maxwell}
\end{equation}
which are invariant under the transformations
\begin{equation}
\eqalign{
F^{+I}_{\mu\nu}&\longrightarrow \tilde F^{+I}_{\mu\nu} =
U^I{}_J\, F^{+J}_{\mu\nu}+Z^{IJ}\,G^+_{\mu\nu J}\,.\cr
G^+_{\mu\nu I}&\longrightarrow \tilde G^+_{\mu\nu I}= V_I{}^J\,
G^+_{\mu\nu J}+W_{IJ}\,F^{+J}_{\mu\nu}\,, \cr }\label{FGdual}
\end{equation}
where $U$, $V$, $W$ and $Z$ are constant, real,  $(n+1)\times(n+1)$
matrices.
The transformations for the anti-selfdual tensors follow by
complex conjugation.
However, to ensure that (\ref{defG}) remains satisfied with
a symmetric tensor $\cal N$, at least in the generic case, the
transformation (\ref{FGdual}) must be symplectic
(disregarding an overall
multiplication of the field strength tensors by a real constant).
More precisely,
\begin{equation}
{\cal O}\ \mathrel{\mathop{=}\limits^{\rm def}}\
\left(\begin{array}{cc} U & Z \\[1mm] W & V \end{array}\right)
\label{uvzwg}
\end{equation}
must be an $Sp(2n+2,{\bf R})$ symplectic matrix, that is, it
must satisfy
\begin{equation}
{\cal O}^{-1} = \Omega\, {\cal O}^{\rm T} \,\Omega^{-1} \qquad
\mbox{where} \quad \Omega =\left(\begin{array}{cc}
0&1 \\[1mm]
-1 & 0
\end{array}\right) .
\label{spc}
\end{equation}
For the sub-matrices $U$, $V$, $W$ and $Z$, this means
\begin{equation}
\eqalign{&U^{\rm T} V- W^{\rm T} Z = V^{\rm T}U - Z^{\rm T}W =
{\bf 1}\,,\cr
&U^{\rm T}W = W^{\rm T}U\,,\qquad Z^{\rm T}V= V^{\rm T}Z\,.\cr}
\label{spc2}
\end{equation}
Incidentally, it follows straightforwardly from
(\ref{FGdual}) that the kinetic
term of the vector fields (\ref{Maxwell}) does not generically
preserve its form under $Sp(2n+2,{\bf R})$ because
\begin{equation}
\eqalign{
\tilde F^{+I}_{\m\n}\,\tilde G_I^{+\m\n} = &F^{+I}_{\m\n}\,
G_I^{+\m\n}  + (U^{\rm T}W)_{IJ} \,F^{+I}_{\m\n}
F^{+\m\n J} \cr
&+ 2  (W^{\rm T}Z)_I{}^J \,F^{+I}_{\m\n}G_J^{+\m\n}
+(Z^{\rm T}V)^{IJ} \, G^+_{\m\n I} G_J^{+\m\n}\,, \cr}\label{acchange}
\end{equation}
which confirms that generally it is only the combined equations of
motion and Bianchi identities that are equivalent, but not the
Lagrangian or the action.

Next, consider the transformation rules for the
scalar fields.
$N=2$ supersymmetry relates the $X^I$ to the field strengths
$F^{+I}_{\m\n}$, while the $F_I$ are related to the $G^{+\m\n}_I$.
Hence, eqs.~(\ref{FGdual}) suggest
\begin{eqnarray}
\tilde X{}^I&=&U^I_{\ J}\,X^J + Z^{IJ}\,
F_J,\nonumber\\
\tilde F{}_I&=&V_I{}^J\,F_J + W_{IJ}\,X^J ,
\label{xxx}
\end{eqnarray}
Owing to the symplectic conditions (\ref{spc}), the quantities
$\tilde F_I$ can be written as the derivative of a new function
$\tilde F(\tilde X)$ with respect to the new coordinate $\tilde
X^I$:
\beq
\tilde F(\tilde X)\
=\ {\textstyle{1\over2}} \big(U^{\rm T}W\big)_{IJ}X^I X^J
+{\textstyle{1\over2}}
\big(U^{\rm T}V + W^{\rm T}  Z\big)_I{}^J X^IF_J
+{\textstyle{1\over2}} \big(Z^{\rm T}V\big){}^{IJ}F_I F_J \,,
\label{Ftilde}
\eeq
where we made use of the homogeneity of $F$.\footnote{%
    For rigid $N=2$ supersymmetry where $F$ is not homogeneous, one
    should add $F(X)-{1\over2}X^I F_I(X)$ to the right-hand side
    of (\ref{Ftilde}) and one may also add terms constant or
    linear in $\tilde X^I$.%
    }
Lagrangians  parameterized by
$F(X)$ and $\tilde F(\tilde X)$ represent equivalent theories, at
least for the Abelian sector of the theory. Furthermore, when
\begin{equation}
\tilde F(\tilde X) = F(\tilde X)\,, \label{Finv}
\end{equation}
the field equations are {\em invariant} under the symplectic
transformations. Note that this does {\em not} imply that $F$
itself is an invariant function in the usual sense. Indeed, from
comparison to
(\ref{Ftilde}) one readily verifies that $F(\tilde X)\not= F(X)$,
as was already observed in \cite{DWVP} for infinitesimal
transformations. A consequence of (\ref{Finv}) is that
substituting $\tilde X$ for $X$ in $F_I(X)$ induces precisely the
symplectic transformation specified in the second
formula of (\ref{xxx}). In practice this is a more direct way to
verify the invariance, rather than checking (\ref{Finv}).

Let us now present some additional details on
the generic transformation rules for various tensors.
In view of
recent interest in duality transformations for rigidly
supersymmetric Yang-Mills theories, we stress that most of these
results apply both to local and rigid $N=2$ supersymmetry. We
have seen that $(X^I,F_J)$ and $(F^{+I}_{\m\n}, G^+_{\m\n J})$
transform linearly as $(2n+2)$-component vectors under
$Sp(2n+2,{\bf R})$. Defining
\begin{equation}
{\partial\tilde X^I\over\partial X^J}\equiv {\cal S}^I{}_{\!J}(X)
= U^I{}_{\!J} +Z^{IK}\,F_{KJ}  \ ,  \label{cSfin}
\end{equation}
we note  that the second derivative of $F(X)$ changes as a
period matrix under $Sp(2n+2,{\bf R})$,
\begin{equation}
\tilde F_{IJ} = \big(V_I{}^KF_{KL}+W_{IL}\big) \,\big[{\cal
S}^{-1} \big]^L{}_{\!J} \,, \label{F2sp}
\end{equation}
where the right-hand side depends on the coordinates $X^I$
through $F_{IJ}$ and $\cal S$. For this reason $(X^I,F_J)$ are
called the periods; in string theory they
correspond to the periods of certain harmonic forms and in
rigidly supersymmetric Yang-Mills theory
they are also very useful
for understanding the non-perturbative features of the theory
\cite{SeibWit}.
Actually, the periods are more fundamental than
the underlying function $F(X)$, because there are situations
where the transformations $\tilde X(X)$ are singular, so that a
meaningful holomorphic prepotential cannot be found; this was demonstrated
recently in \cite{Ceresole}.
Although in this case it is still possible to define a
homogeneous function $\tilde F(\tilde X)={1\over 2}\tilde
X{}^I\tilde F_I$, this function does not depend on all the
coordinates $\tilde X^I$.
Often, such a $\tilde F(\tilde X)$ vanishes identically, but this
is not generic.
In spite of the non-existent prepotential, however,
the underlying theory is well defined and can be related via symplectic
reparameterizations to theories where a prepotential does exist
\cite{Ceresole}.
Such situations are indicated by a singular matrix of second
derivatives $\tilde F_{IJ}$ (cf.~(\ref{F2sp})) while the matrix
$\tilde{\cal N}_{IJ}$ of inverse gauge couplings remains well-behaved
(cf.~(\ref{nchange}) below).
In the purely Abelian case, one can always choose a coordinate basis
$X^I$ for which the prepotential $F$ does exist, but this may
fail in the presence of a non-Abelian gauge group where
electric-magnetic dualities are not legitimate.
In section 4, we shall see that in string theory, the uniform
dilaton dependence of the gauge couplings is manifest only in a
basis where $F$ does not exist,
although this is not an obstacle for considering the loop corrections
to the prepotential defined for another basis.
Here, we would like to add a comment that in situations without
a prepotential, the appropriate criterion for invariance of the
equations of motion is not eq.~(\ref{Finv}) (which is not quite
 meaningful in this case), but whether
the transformation of the periods can be correctly induced (up to
an overall holomorphic proportionality factor) by
appropriate changes of the underlying coordinates $z^A$.

The transformation rules for the tensors ${\rm Im}(F_{IJ})$ and
$F_{IJK}$ are as follows:
\beq
\eqalign{
{\rm Im}(\tilde F_{IJ})\ &
=\ {\rm Im}(F_{KL})\, \big[\bar{\cal S}^{-1}\big]^K{}_{\!I}\,
    \big[{\cal S}^{-1}\big]^L{}_{\!J}\, \cr
\tilde F_{IJK}\ &
=\ F_{MNP}\, \big[{\cal S}^{-1}\big]^M{}_{\!I} \,
    \big[{\cal S}^{-1}\big]^N{}_{\!J}\,
    \big[{\cal S}^{-1}\big]^P{}_{\!K} \,.
}\label{Nsp}
\eeq
The transformation rules for the gauginos are also given by the
${\cal S}^{-1}$ or its complex conjugate, depending on chirality.
In the rigidly supersymmetric case, the transformation of
the gauge  field strengths $F^{\pm I}_{\m\n}$
is also described by this matrix, but in the locally
supersymmetric case this transformation is modified as the
tensor ${\cal N}_{IJ}$ governing the relation between
the  $F^{\pm I}_{\m\n}$ and the $G^{\pm}_{\m\n I}$ is no longer equal to
${1\over 4}\bar F_{IJ}$ or even anti-holomorphic.
Nevertheless,
the transformation rule for the tensor $\cal N$ itself is
precisely as in the rigid case, namely
\begin{equation}
\tilde {\cal N}_{IJ} = (V_I{}^K {\cal N}_{KL}+ W_{IL} )\,
\big[(U+ Z{\cal N})^{-1}\big]^L{}_J  \,.\label{nchange}
\end{equation}
To obtain this last result it is crucial that the function $F(X)$ is
homogeneous of second degree. Note that the symmetry of $\tilde
F_{IJ}$ and $\tilde{\cal N}_{IJ}$ is ensured by
the symplectic conditions (\ref{spc}).

Three particular subgroups of the
$Sp(2n+2,{\bf R})$ will be relevant to our discussion
in section~4.
The first subgroup contains the classical
target-space duality transformations which are symmetries of the
tree-level Lagrangian. From eqs.~(\ref{acchange}), (\ref{xxx})
we learn that the Lagrangian is left invariant by the subgroup
that satisfies $W=Z=0$ and $V^T=U^{-1}$.
For the second subgroup, we continue to demand $Z=0$ but relax
the $W=0$ condition; according to eq.~(\ref{spc2}), we then should have
$V^{\rm T}=U^{-1}$ and $W^{\rm T} U$ should be a symmetric matrix.
These conditions lead to {\em semiclassical} transformations
of the form
\begin{equation}
\eqalign{
\tilde X^I  &=U^I{}_J\,X^J\,, \cr
\tilde F^{\pm I}_{\m\n} &= U^I{}_J\,F^{\pm J}_{\m\n}\,,  \cr}
\qquad
\eqalign{
\tilde F_I &=  [U^{-1}]^J{}_I\, F_J  +  W_{IJ} X^J\, , \cr
\tilde{\cal N} &= \big[U^{-1}\big]^{\rm T}{\cal N}U^{-1} + W
U^{-1} \,,\cr} \label{Tsymp}
\end{equation}
which can always be implemented as Lagrangian symmetries of the
vector fields $A^I_\mu$.
The last term in the last equation in (\ref{Tsymp}) amounts to
a constant shift of the theta angles; at the quantum level,
such shifts are quantized and hence the
symplectic group must be restricted to $Sp(2n+2,{\bf Z})$.
We will see that such shifts in the $\theta$-angle do occur
whenever  the one-loop gauge couplings have  logarithmic singularities
at special points in the moduli space where massive modes
become massless. Therefore, these symmetries are related to the
semi-classical (one-loop) monodromies around
such singular points.
The third subgroup contains elements that interchange the
field-strength tensors $F^I_{\m\n}$ and $G_{\m\n I}$ and
correspond to electric-magnetic dualities. These
transformations are defined by
$U=V=0$ and $W^{\rm T}=- Z^{-1}$, which yields
\begin{equation}
\tilde{\cal N} = -  W \,{\cal N}^{-1}\,W^{\rm
T}\,,  \label{Ssymp}
\end{equation}
so that they give rise to an inversion of the gauge couplings and
hence must be non-perturbative in nature.
In the heterotic string context, such transformations are often
called $S$-dualities because of the way they act upon the dilaton
field $S$.
We shall return to this issue in section 4.1.

Let us now turn our attention to the holomorphic prepotentials
of the following special form
\begin{equation}
F(X)={\textstyle{1\over2}}d_{ABC}{X^AX^BX^C\over X^0}\,,\label{dfunction}
\end{equation}
where $d_{ABC}$ are some real constants.
The theories described by this class
of $F$-functions emerge via dimensional reduction from
five-dimensional $N=2$ supergravity coupled to vector
multiplets;
they also emerge in the heterotic string context, regardless
of any dimensional reduction.
Let us therefore record a few convenient formulae for future use.
In special coordinates,
\begin{eqnarray}
{\cal W}_{ABC} &=& 3d_{ABC}\,,\nonumber  \\
{\rm Im}(F_{AK})X^K(z)&=&-{\textstyle{3i\over 4}} d_{ABC}\,
(z-\bar z)^B(z-\bar z)^C\,,\nonumber  \\
{\rm Im}(F_{0K})X^K(z)&=& {\textstyle{i\over 4}} d_{ABC}\big(z^A z^B z^C -
3\bar z^A \bar z^B z^C + 2\bar z^A \bar z^B \bar z^C \big)\,,
\nonumber \\
{\rm Im}(F_{KL})\,X^K(z)\,\bar X^L(\bar z)&=&-{\textstyle{1\over
    2}}e^{-K(z,\bar z)} \,,\nonumber \\
{\rm Im}(F_{KL})\, X^K(z)\,X^L(z)&=& e^{-K(z,\bar z)} \,,
\end{eqnarray}
while the K\"ahler potential is given by
\begin{equation}
K(z,\bar z)=-\log \Big(-{\textstyle{1\over 2}}id_{ABC}\,(z-\bar
z)^A(z-\bar z)^B(z-\bar z)^C\Big)  \,.
\end{equation}
The special K\"ahler spaces corresponding to (\ref{dfunction})
always possess continuous isometries \cite{BEC}, which in special
coordinates take the form \cite{DWVVP}
\begin{equation}
\delta z^A = b^A - \textstyle{2\over 3} \beta \,z^A +\tilde
B^A_{\;B}\, z^B - \textstyle{1\over 2} \big(R^A{}_{\!BC}{}^{\!D}\,
a_D\big)\, z^B z^C .
\label{zisometry}
\end{equation}
Here $\b$, $b^A$ and $a_A$ are real parameters and the matrix $\tilde B$
parameterizes the infinitesimal real linear transformations of the
$X^A$ under which $d_{ABC}\,X^AX^BX^C$ is left invariant; the
isometries corresponding to
the parameters $a_A$ exist only for those parameters for which
the $R^A{}_{\!BC}{}^{\!D}\,a_D$ are constant.
All homogeneous spaces of this type have been classified and it
has been shown that all their isometries are related to the symplectic
transformations discussed above \cite{BEC,DWVVP}.
The infinitesimal form of the
matrices $U$, $V$, $W$ and $Z$ was first determined in
\cite{BEC}. Introducing the notation
\begin{equation}
{\cal O} ={\bf 1} + \left(\begin{array}{cc}
B & -D \\[2mm]
C &-B^{\rm T}
\end{array}\right) , \label{Oinf}
\end{equation}
we have (the first row and column refer
to the $I=0$ component)
\begin{equation}
B^I_{\,J} =\left(\begin{array}{cc}
\beta & a_B \\[2mm]
 b^A &\tilde B^A_{\;B} +\frac{1}{3}\beta \,\delta^A_{\,B}
\end{array}\right) ,
\ \
C_{IJ} =\left(\begin{array}{cc}
0 & 0 \\[2mm]
 0 & 3 (d\,b)_{AB}
\end{array}\right) ,
\ \
D^{IJ} =\left(\begin{array}{cc}
 0 & 0 \\[2mm]
 0 & -\frac{4}{9} (C\,a)^{AB}
\end{array}\right) .    \label{matrixduald}
\end{equation}
where we use an obvious notation where $(d\,b)_{AB}= d_{ABC}b^C$,
$(d\,bb)_A=d_{ABC}b^Bb^C$, $(d\,bbb)= d_{ABC}b^Ab^Bb^C$, and
likewise for contractions of $C^{ABC}$ with the parameters $a_A$,
where $C^{ABC}$ is defined by
\begin{equation}
C^{ABC} = - \textstyle{\frac{9}{4}}  \,e^{2K(z,\bar z)}\,\bar
{\cal W}{}^{ABC}(z,\bar z)\,.
\label{defCABC}
\end{equation}
Obviously, in general $C^{ABC}$ are not constant , but again the
possible parameters $a_A$ are restricted by the condition that
$C^{ABC}a_C$ should be constant. For homogeneous spaces, there are
always nontrivial solutions for the $a_A$, while for symmetric spaces
$C^{ABC}$ are constant and so there are precisely $n$
isometries associated with $n$ independent parameters $a_A$.
All homogeneous spaces corresponding to the
functions (\ref{dfunction}) have been classified in \cite{DWVVP}.

In string theory the continuous isometries are not preserved by
world-sheet instanton effects. Therefore infinitesimal
isometries and corresponding duality transformations are only
relevant for certain couplings. In that context we observe that
the isometries associated with the parameters $b^A$
and $a_A$ can simply be exponentiated to finite symplectic
transformations, which
for special values of $b^A$ and $a_A$ are exact symmetries
of the underlying string theory.
Of course, these symmetries can also be determined from
string arguments alone and it will be instructive to compare the
results. For the symplectic transformation associated with finite
parameters $b^A$, we have
\begin{equation}
\eqalign{
U(b)= V^{\rm T}(-b)=  \pmatrix{1&0\cr
\noalign{\vskip 2mm}
               b^A&{\bf 1}_n\cr} \,,\quad
W(b)= {\textstyle{1\over 2}}\pmatrix{-(d\,bbb)& - 3(d\,bb)_B\cr
\noalign{\vskip 3mm}
               3(d\,bb)_A& 6 (d\,b)_{AB}\cr} \,,\quad
Z(b)=0\,,\cr} \label{btransform}
\end{equation}
with the corresponding transformations of the special coordinates
being simply
\begin{equation}
z^A\longrightarrow z^A + b^A  \,.
\end{equation}
These transformations are of type (\ref{Tsymp} and thus can be
realized on the vector potentials and leave the
Lagrangian invariant up to a total divergence corresponding to a
shift in the $\theta$ angles.%
\footnote{%
     We should stress that this result holds in the basis
     associated with (\ref{dfunction}). As we shall exhibit in
     subsection~4.1, in another symplectic basis the situation
     can be qualitatively different.
     }
For specific choices of the
$d_{ABC}$, the above  results can now be compared to those
derived directly from string theory \cite{FLT,Shevitz,FKLZ}.

Likewise, the symplectic transformations associated with finite
parameters $a_A$ are
\begin{equation}
\eqalign{
U(a)= V^{\rm T}(-a)=  \pmatrix{1&a_B\cr
      \noalign{\vskip 2mm}
               0&{\bf 1}_n\cr} \,,\quad
W(a)=0\,,\quad
Z(a)= {\textstyle{2\over 27}}\pmatrix{(C\,aaa)& 3(C\,aa)^B\cr
         \noalign{\vskip 3mm}
               -3(C\,aa)^A& 6 (C\,a)^{AB}\cr}
\,,\cr} \label{atransform}
\end{equation}
and the corresponding transformations on special coordinates take
the form
\begin{equation}
z^A\longrightarrow {z^A +{2\over 3}(C\,a)^{AB}(d\,zz)_B +{1\over
9}(C\,aa)^A(d\,zzz)  \over 1 +a_Bz^B +{1\over 3}(C\,aa)^B (d\,
zz)_B - {1\over 27}(C\,aaa)(d\,zzz) }\,. \label{aztransform}
\end{equation}
For appropriate values of the parameters the matrices
(\ref{btransform}) and (\ref{atransform}) may generate
the group of discrete transformations that are preserved at the
quantum level. For the symmetric K\"ahler space relevant for the
heterotic string compactifications, the $SL(2,{\bf Z})$ groups
associated with target space- and $S$-dualities can be generated in
this way, as their nilpotent subgroups are special cases of
(\ref{btransform}) and (\ref{atransform}). We return to this in
section~4.1.

\subsect{Quantum effects in $N=2$ theories}

Classically, the geometry of the field space is unrelated to the
field dependence of the particles' masses.
However, an effective quantum field theory (EQFT) has to be cut-off at
the Planck scale and thus should not include any of the superheavy states.
The distinction between the light fields that should be  manifest in the
low-energy EQFT and the heavy fields that should  be integrated out
depends on the moduli parameters of the underlying string vacuum.
In general, a single connected family of string vacua gives rise to several
distinct low-energy EQFTs according to the moduli-dependent spectra
of the light particles.
Therefore, from any particular EQFT's point of view, there is a difference
between the spectrum-preserving moduli scalars, whose vacuum expectation
values  may become
arbitrarily large without giving a Planck-sized mass to any otherwise
light particles, and between all the other flat directions of the scalar
potential.
Physically, the latter may also develop arbitrarily large
vacuum expectation values, but in that
case the spectrum of the light particles would no longer agree with the
original EQFT and one has to switch over to a different EQFT in order to
properly describe the low-energy limit of the string vacuum.
Consequently,  the field-dependent couplings of the  EQFT should
be written as complete analytic functions of the
spectrum-preserving moduli fields, but their dependence on all
the other field may be described by a truncated power series.
(See ref.~\cite{KLb} for a more detailed discussion.)

In this spirit, we
divide the scalars $z^A=X^A/X^0$ belonging to
vector multiplets of an $N=2$ locally supersymmetric EQFT into
the spectrum-preserving moduli $\mod^\a=-iz^\a$ and the ``matter'' scalars
$\mat^a=-iz^a$\footnote{%
    The $-i$ is included
    in order to be consistent with the standard string conventions.
    For toroidal compactifications,
    $\mod^\a$ correspond to the $3+P$ complex moduli fields
    $S$, $T$, $U$ and $\phi^i$ introduced in section~2.
    }
and expand the prepotential $\Fc$ of the theory as a truncated power
series in the latter:
\beq
\Fc(\mod,\mat) \ =\ \fone(\mod)\
+\sum_{ab} \ftwo_{ab}(\mod)\mat^a\mat^b\ +\ \cdots .
\label{Fexp}
\eeq
Obviously, all scalars in the non-Abelian vector multiplets should
be regarded as matter (there are flat directions among these
scalars, but none are spectrum preserving since their
vacuum expectation values induce a
mass for some of the non-Abelian fields);
for such non-Abelian matter, the gauge symmetry of the prepotential
requires
\beq
\ftwo_{ab}(\mod)\ =\ \delta_{ab} \ftwo_\ind(\mod)
\label{NonAbelianFA}
\eeq
where the index $\ind$ refers to the appropriate irreducible factor
$G_\ind$ of the gauge group $G=\prod_\ind G_\ind$.
Similarly, if any hypermultiplets appearing in the EQFT are charged
under an Abelian gauge symmetry, the scalar superpartner of that
gauge boson should be regarded as matter since its
vacuum expectation value would give masses
to all such charged hypermultiplets.
On the other hand, if all the {\em light} particles are neutral with
respect to some Abelian gauge field, then its scalar superpartner is
a spectrum-preserving modulus.
To be precise, we divide the Abelian vector multiplets into $\mod^\a$
and $\mat^a$ such that all the light hypermultiplets of the EQFT
under consideration
are exactly massless for $\mat^a=0$ and arbitrary $\mod^\a$.

A proper discussion of the field-dependent couplings of an
effective {\em quantum} field theory must distinguish between
two kinds of renormalized couplings~\cite{SV}:
First, there are {\em effective couplings} associated with
physical processes; for example, Coulomb-like scattering
of charged particles defines a momentum-dependent gauge
coupling $g(p^2)$.
The momentum dependence of such couplings is unavoidable
in theories with massless charged particles;
therefore, the effective couplings generally cannot be
summarized in any {\em local} effective Lagrangian.
Second, there are the {\em Wilsonian couplings}, which are the
coefficients of the quantum operator products in the action
functional of the theory.
Similar to its classical counterpart,
this Wilsonian action is an $\int d^4x$ of a local
Wilsonian Lagrangian;
consequently, the supersymmetric constraints satisfied by the Wilsonian
couplings of an EQFT are the same as in the classical case.%
\footnote{%
    This presumes that the quantum theory is regularized in a way
    that preserves both the local supersymmetry and the four-dimensional
    background gauge invariance.
    See ref.~\cite{KLa} for the discussion of these issues in the
    local $N=1$ case.%
    }
In particular, the Wilsonian prepotential of an $N=2$ supersymmetric
EQFT must be a holomorphic function $\Fc(\mod,\mat)$ defining the
Wilsonian K\"ahler function $K(z,\bar z)$ according to
eq.~(\ref{Kspecial}).
In light of the expansion (\ref{Fexp}), we have
\beq
\eqalign{
K(\mod,\modb, \mat, \bar{\mat})\ &
=\  \Khat(\mod,\modb)\
    + \sum_{ab} Z_{ab} (\mod, \modb)\, \mat^a \matb^b\
    +\ \cdots \,,\cr
{\rm where}\quad \Khat(\mod,\modb)\ &
=\ -\log\left[2(\fone + \bar{\fone})\,
     -\sum_\a(\mod^\a+\modb^\a)(\del_\a{\fone}+\bar\del_\a {\fone})
    \right]\ ,\cr
{\rm and}\quad Z_{ab}(\mod,\modb)\ &
=\ 4e^{\Khat(\mod, \modb)}\,{\rm Re}\, \ftwo_{ab}(\mod) .\cr
}\label{Zexp}
\eeq
Similarly, the Wilsonian gauge couplings follow from
eq.~(\ref{Ndef}).
Since the distinction between the matter scalars $\mat^a$ and the
spectrum-preserving moduli $\mod^\a$ (henceforth called simply moduli)
presumes $|\mat^a|\ll1$ (in Planck units), it follows that the
vector superpartners of the $\mat^a$ do not mix with the graviphoton
and hence the corresponding Wilsonian gauge couplings are simply
$(g^{-2}_{ab})^{\rm W}={\rm Re} \ftwo_{ab}(\mod)$.
In particular, for the non-Abelian gauge fields
$(g^{-2}_\ind)^{\rm W}={\rm Re}\,\ftwo_\ind(\mod)$.
On the other hand, the vector superpartners of the moduli $\mod^\a$
do mix with the graviphoton;
consequently, the corresponding Wilsonian gauge couplings
$(g^{-2}_{\a\b})^{\rm W}$, $(g^{-2}_{\a0})^{\rm W}$ and
$(g^{-2}_{00})^{\rm W}$ are
complicated non-holomorphic function of the moduli
in accordance with eq.~(\ref{Ndef}).

In supersymmetric EQFTs, holomorphic quantities are associated
with chiral superspace integrals and consequently enjoy many
no-renormalization theorems.
In particular, in $N=2$ supersymmetric theories, the entire
prepotential $\Fc$ is not renormalized in any higher-loop order
of the perturbation theory \cite{NRT,NS};
in section~4.2  we present an independent
argument for this no-renormalization theorem in the heterotic
string context.
Thus,
\beq
\Fc\ =\ \Fctree\ +\ \Fcol\ +\ \Fc^{(NP)} ,
\label{Floop}
\eeq
where $\Fctree$ is the tree-level prepotential, $\Fcol$
originates at the one-loop level of the EQFT while $\Fc^{(NP)}$
is due to instantons and other non-perturbative effects
(see refs.~\cite{SeibWit,LKTY} for a detailed analysis of such
effects in rigid $N=2$ SSYM.);
in this article we confine our attention
to the purely perturbative properties of
the prepotential $\Fc$ and therefore drop the
$\Fc^{(NP)}$ term from our further discussion.
In terms of the expansion~(\ref{Fexp}), eq.~(\ref{Floop})
(without $\Fc^{(NP)}$) means
\beq
\eqalign{
\fone(\mod)\ &
=\ \fonetree(\mod)\ +\ \foneol(\mod)\ ,\cr
\ftwo_{ab}(\mod)\ &
=\ \ftwotree_{ab}(\mod)\ +\ \ftwool_{ab}(\mod)\  ;\cr
}\label{fhexp}
\eeq
in particular, for the non-Abelian gauge group factors $G_\ind$,
the Wilsonian gauge couplings are
\beq
(g^{-2}_\ind)^{\rm W}\ =\  {\rm Re}\ \ftwotree_\ind(\mod)\
+\ {\rm Re}\ \ftwool_\ind(\mod)\ ,
\label{gna}
\eeq
in complete analogy with the $N=1$ EQFTs.

Thus far we discussed the analytic properties of the Wilsonian couplings.
Let us now turn to  the physical, momentum-dependent effective
gauge couplings $g^{-2}_\ind (p^2)$ which account for
all the quantum effects, both high-energy and low-energy.
As argued in refs.~\cite{DKL,AGN,KLa,HOLAN,KLb},
the low-energy effects due to light charged
particles give rise to a non-holomorphic moduli dependence
of these effective gauge couplings;
a supersymmetric Ward identity
$\del_\mod g^{-2}_\ind = i{} \theta_{\ind,\mod}$
relates this non-holomorphicity to the non-integrability of the
effective axionic couplings
$ \theta_{\ind,\mod} \neq\del_\mod \theta_\ind(\mod)$.
According to this Ward identity, the entire moduli dependence of the
effective gauge couplings can be derived from that of the axionic couplings,
which in turn follows from the connection terms proportional
 to $\del_\mu\mod^\a$
in the  Lagrangian for the charged fermions of the theory.
For the moduli $\mod^\a$ belonging to $N=2$ vector multiplets,
these connection terms have exactly the same form as in  local $N=1$
supersymmetry, so that we may simply adapt the
$N=1$ formula of \cite{KLa} to our present case.
Thus, we find that to all orders in perturbation theory
\cite{SV,KLa,HOLAN,KLb},\footnote{%
    In our notations, $n_r$ is the number of charged hypermultiplets in the
    representation $r$ of the gauge group,
    $T_\ind(r) \delta^{a b}={\rm Tr}_r( T^a T^b )$ ($T^a$ being the hermitian
    generators of the gauge group $G_\ind$),
    $T(G_\ind)$ abbreviates $T_\ind({\rm adjoint~of~} G_\ind)$,
    and $b_\ind=2 \sum_r n_r T_\ind(r)-2 T(G_\ind)$
    is the beta-function coefficient of the $N=2$ gauge theory.
    We presume $p\ll\mpl$.
    }
\begin{equation}
\eqalign{
g^{-2}_\ind (\mod, \modb, p^2)\ &
=\  {\rm Re}\, f_\ind (\mod)\ +\ {b_\ind\over 16\pi^2}\,\log{\mpl^2\over p^2}\
+\ { \sum_r n_r T_\ind (r) \over 8\pi^2}\,
\Khat (\mod, \modb)\cr
&+\ {T (G_\ind)\over 8\pi^2}\,\log g^{-2}_\ind(\mod, \modb, p^2)\
-\ {T(G_\ind)\over 8\pi^2}\,\log Z_\ind (\mod, \modb, p^2)\
+\ \rm const,
}\label{running}
\end{equation}
where $Z_\ind$ is the effective normalization factor for the scalar
superpartners of the gauge bosons of $G_\ind$ and we make use
of an $N=2$ supersymmetric Ward identity that prevents similar normalization
matrices for the hypermultiplets from depending on the moduli belonging
to vector multiplets.
Furthermore, $N=2$ supergravity provides for
\beq
Z_\ind (\mod, \modb, p^2)\
= 4 e^{\Khat(\mod, \modb)}\, g^{-2}_\ind(\mod, \modb, p^2);
\label{Zrel}
\eeq
this relation is also valid to all orders of the perturbation theory
as long as $\Khat$ is derived from the quantum-corrected $h(\mod)$ rather
than the tree-level $h^{(0)}(\mod)$.
Substituting (\ref{Zrel})  into (\ref{running}), we arrive at
\beq
g^{-2}_\ind(\mod, \modb, p^2)\
=\ \mathop{\rm Re} f_\ind (\mod)\
+\ {b_\ind\over 16\pi^2}\,
\Big(\log{\mpl^2\over p^2} +  \Khat(\mod, \modb)\Big)\
+\ \rm const
\label{anofin}
\eeq
(for the Abelian effective gauge couplings $g^{-2}_{ab}(\mod,\modb,p^2)$,
read $b_{ab}=2\sum_r n_r {\rm Tr}_r( T_a T_b )$ for the $b_\ind$).
Although the above argument might suggest that the ``constant''
term in this formula could be a function of the moduli belonging
to hypermultiplets,
actually the effective gauge couplings are completely independent of any
hypermultiplet moduli.
For consistency's sake, we have verified this statement by explicitly
calculating the axionic couplings, but it can be better understood as
a Ward identity of the $N=2$ supersymmetry, rigid or local:
The $N=2$ gauge fields couple to {\sl charged} hypermultiplets
in a minimal gauge-covariant way
(and hence the spectrum of such hypermultiplets affects the
beta-function coefficients $b_\ind$ in (\ref{anofin})),
but they do not have two-derivative couplings to any {\sl neutral}
hypermultiplets
and hence the gauge couplings cannot depend on the latter.

In $N=1$ supersymmetric gauge theories, the scalar normalization factors
on the right hand side of eqs.~(\ref{running}) renormalize differently
from those of the gauge fields;
this leads to the higher-loop renormalization of
the effective gauge couplings even though the Wilsonian gauge couplings
renormalize only at the one-loop level \cite{SV}
(or non-perturbatively).
The extended $N=2$ supersymmetry eliminates this effect and hence
the perturbative renormalization of both
the Wilsonian and the effective gauge couplings stops at the one-loop level.
Indeed, in the rigid case, the only difference between the two kinds of
gauge couplings is the moduli-independent
$(b_\ind/16\pi^2)\log \mpl^2/p^2$ term
and the moduli dependence of both couplings can be described by the same
holomorphic function $f_\ind(\mod)$;
this behavior was important for the non-perturbative analysis
of refs.~\cite{NS,SeibWit}.

However, for the {\sl locally} $N=2$ supersymmetric gauge theories,
(\ref{anofin}) tells us that although the effective gauge couplings
do not renormalize at higher-loop orders of the perturbation theory, their
moduli dependence is different from that of the corresponding Wilsonian
couplings (unless $b_\ind=0$):
In addition to a harmonic function ${\rm Re}\,f_\ind(\mod)$, the effective
gauge coupling also contains a K\"ahler term.
Note that this is exactly the behavior observed in the explicit string-loop
calculation of the gauge couplings of
the toroidal compactifications of $(d=6,N=1)$ vacua in ref.~\cite{DKL},
where the non-harmonic term in the string-threshold correction to the
effective $g_\ind^{-2}$ was found to be precisely $b_\ind$ times
the K\"ahler function of the toroidal moduli.

We conclude this section with a discussion of
special points or subspaces of the moduli space
where otherwise heavy particles become massless.
For the sake of definiteness, let us assume that $\phi$ is a
spectrum-preserving modulus of the EQFT as long as $|\phi|\not\ll1$
but for $\phi=0$ the gauge group becomes enlarged because of additional
massless vector multiplets, although the case of additional charged
hypermultiplets can be handled in exactly the same way.
Clearly, string vacua corresponding to $|\langle\phi\rangle|\ll1$
have to be described by a different EQFT and in that new EQFT the
$\phi$ scalar itself is no longer a spectrum-preserving modulus but
a matter scalar $\mat^a$ (or perhaps a linear combination of such $\mat^a$).
However, in the range of {\em moderately} small $|\langle\phi\rangle|$,
both EQFTs are valid and should yield identical low-energy physical
quantities.
Therefore, we can use this overlap of the two EQFTs' domains of validity
to relate their Wilsonian couplings to each other.

Physically, moderately small $|\langle\phi\rangle|$ means that there is
a threshold at the energy scale $M_I\sim |\langle\phi\rangle|\mpl$ that
is well below the Planck scale but well above the scale one uses to measure
the {\em low}-energy physical quantities.
In this range, the difference between the small-$\phi$ EQFT
and the large-$\phi$ EQFT is that the fields with $O(M_I)$ masses
are present  in the former but are integrated out from the latter.
Therefore, the difference between the Wilsonian gauge couplings of
the two EQFTs is simply a threshold correction.
Ref.~\cite{KLa} gives a formula for such threshold corrections for $N=1$
supersymmetric gauge theories and it is easy to see that it applies
without any modifications in the present $N=2$ case.%
\footnote{%
    Actually, the $N=2$ case is simpler because the masses of the
    short vector multiplets or hypermultiplets have to saturate
    the Bogomolny\u\i\ bound.
    (The long vector multiplets' net contribution to beta-functions is
    zero and hence they do not contribute to the the threshold corrections
    either.)
    The threshold corrections to the Wilsonian gauge couplings involve
    the {\em unnormalized} masses of these multiplets \cite{KLa},
    which are simply proportional to $\phi\mpl$ with coefficients that
    do not depend on any other moduli.
    This explains why the constant term in eq.~(\ref{DFA}) is indeed
    constant.%
    }
Thus, to all orders of the perturbation theory,
\beq
\ftwo'_\ind\ -\ \ftwo_\ind\
=\ {b'_\ind-b_\ind\over16\pi^2}\,\log\phi^2\ +\ {\rm const},
\label{DFA}
\eeq
where the primed quantities refer to the large-$\phi$ EQFT and the
unprimed to the small-$\phi$ EQFT.
Modulo a trivial change of notations, this relation also holds for
the Abelian gauge couplings of the two EQFTs, including the gauge
couplings of the vector member of the $\phi$ supermultiplet itself.
Since in the $|\phi|\ll1$ limit, the Wilsonian gauge coupling
$(g^{-2}_{\phi\phi})^{\rm W}$ of the large-$\phi$ EQFT
is simply ${1\over2}{\rm Re}\,\del_\phi^2\fone'$,
it follows that the $\phi$ dependence of the moduli prepotential
$\fone'$ of that EQFT must have the form
\beq
\fone'(\phi,\mod)\ =\ \fone(\mod)\
- \ {b_{(\phi)}\over16\pi^2}\left[ \log\phi^2 +{\rm const}\right]\phi^2\
+\ \ftwo_{(\phi)}(\mod)\phi^2\
+\ O(\phi^3),
\label{FONEphi}
\eeq
where $b_{(\phi)}$ is the $\beta$-function coefficient of the
gauge group under which $\phi$ is charged at $\phi=0$,
$f_{(\phi)}$ is the corresponding Wilsonian coupling and
the $O(\phi^3)$ term is completely regular in the $\phi\to0$ limit
({\it i.e.}, it is a convergent power series
$A_3(\mod)\phi^3+A_4(\mod)\phi^4+\cdots$).
Later in this article (section~4.3), eq.~(\ref{FONEphi}) will help
us to completely determine the one-loop moduli prepotential
$\foneol(T,U)$ for toroidal compactifications of the six-dimensional
heterotic string.

%
\sect{Low-energy $N=2$ effective theories for Heterotic String Vacua}
\setcounter{footnote}{0}

In the previous section we reviewed
the  couplings of $N=2$ vector multiplets at the classical and
quantum level.
In this section we study the effective Lagrangian of
$N=2$ heterotic vacuum families and display the special properties
of the  prepotential $F$ which arise in these theories.

\subsect{Classical results}

As we discussed in section~2, the dilaton and the antisymmetric
tensor gauge field in $N=2$ heterotic compactifications are
accompanied by an Abelian vector gauge field. Together they are
contained in a new $N=2$ supermultiplet, called the vector-tensor
multiplet, which is dual to an Abelian vector multiplet. The
scalar component $S$ of the latter includes the dilaton as its
real part and the axion as its imaginary part.
The couplings of the dilaton multiplet are independent of the
properties of the internal SCFT and thus
universal at the string tree level; in particular, the dilaton
does not mix with any of the other scalar fields in the spectrum
of the EQFT. Furthermore, the axion is subject to a continuous
Peccei-Quinn symmetry, which implies that the K\"ahler
potential is only a function of $(S+\bar S)$.
Both properties  together imply that the moduli space
contains the dilaton field $S$ as the complex coordinate of a
separate $SU(1,1)/U(1)$ factor. The only special K\"ahler manifold
of any dimension $n>1$
that  satisfies this constraint is the  symmetric space \cite{FVP}
\begin{equation}
{SU(1,1)\over U(1)}\otimes {SO(2,n-1)\over SO(2)\times
  SO(n-1)}\, , \label{SKP}
\end{equation}
with a  prepotential  (up to symplectic reparametrizations)
\begin{equation}
F(X)= -{X^1\over X^0}\Big[X^2X^3-\sum_{I = 4}^n(X^{I})^2\Big]\,.
\label{ourF}
\end{equation}
The moduli $\mod$ of the previous section are  identified
with
\begin{equation}
S=-i {X^1\over X^0}\,,\quad T=-i {X^2\over X^0}\,,\quad
U=-i {X^3\over X^0}\,,\quad \phi^i=-i {X^{i+3}\over X^0}\ ,\quad
(i=1, \ldots, P)\ ,
\label{modid}
\end{equation}
while the remaining $X^I$ correspond to
non-moduli scalars $\mat^a = -i X^{a+P+3}/ X^0$
($a = 1, \ldots, (n=P-3)$).
$T$ and $U$ can be thought of as the toroidal moduli introduced
in section~2. However, even in non-toroidal string vacua
the prepotential $F$ is given by (\ref{ourF}), where
$X^2$ and $X^3$ can be any two moduli.\footnote{%
    If the heterotic vacuum's spectrum contains only one
    modulus in addition to $S$, that modulus should be identified
    with $(T+U)/2$ while the diffference $(T-U)/2$ is frozen
    at zero value.
    In terms of eq.~(\ref{ourF}), this means simply treating
    $X^2=X^3$ as a single independent coordinate.
    On the other hand, when the vacuum has no moduli at all besides
    $S$, the prepotential (\ref{ourF}) is incompatible with
    non-zero gauge couplings for the non-moduli gauge fields and
    one {\em must} use coordinates for which $F$ does not exist.
    However, we shall see momentarily that such coordinates are
    convenient for all heterotic $N=2$ vacua.%
    }
The moduli space (\ref{SKP}) has been analyzed
in detail in refs.~\cite{SFetal,Ceresole}
The K\"ahler potential is easily calculated from
eqs.~(\ref{KP}), (\ref{ourF}), (\ref{modid}) to be
\begin{equation}
K= -\log \Big( (S+\bar S) \Big[(T+\bar T)(U+\bar U)-\sum_{i}
(\phi^i+\bar\phi^i)^2 -\sum_{a}(\mat^a+\bar\mat^a)^2 \Big]\Big)\, .
\label{Ktree}
\end{equation}
In terms of our previous notation (eq.~(\ref{Fexp})--(\ref{gna}))
we have
\beq
\eqalign{
\fonetree\ &=\ - S\Bigl(TU- \sum_i(\phi^i)^2\Bigr) \,,\qquad
\ftwotree\ =\ S\, ,\cr
\Khat\ &=\ -\log (S+\bar S) -\log\Bigl[(T+\bar T)(U+\bar U)-\sum_{i}
	(\phi^i+\bar\phi^i)^2\Bigr] , \cr
Z\ &=\ {2\over (T+\bar T)(U+\bar U)-\sum_{i} (\phi^i+\bar\phi^i)^2}\,.\cr
}\label{treecouplings}
\eeq
In particular, for any the non-Abelian factor  in
the gauge group $G$ (or more generally any non-moduli vector multiplets)
the tree-level gauge coupling is universal and depends only on
the dilaton's vacuum expectation value,
\beq
g^{-2}_\ind = {\rm Re}\, S \qquad  {\rm for\  all}\   \ind ,
\eeq
which is indeed a well-known tree-level property of the heterotic string.

On the other hand, the gauge couplings for for the vector superpartners
of the moduli scalars are given by the non-holomorphic matrix
${\cal N}_{IJ}$.
Substituting eq.~(\ref{ourF}) into eq.~(\ref{Ndef}),
we find
\begin{equation}
\eqalign{
{\cal N}_{TT}&
=-{\textstyle{i\over 2}} e^\Khat\, (S+\bar S)^2(U+\bar U)^2\,, \qquad
{\cal N}_{UU}
=-{\textstyle{i\over 2}} e^\Khat\, (S+\bar S)^2(T+\bar T)^2\,,  \cr
{\cal N}_{TU}&
=i\bar S-{\textstyle{i\over 2}} e^\Khat(S+\bar S)^2(T+\bar T)(U+\bar U)\,,\cr
{\cal N}_{ij}&
= -2i \bar S\, \delta_{ij} -2ie^\Khat\, (S+\bar S)^2
	(\phi^i+\bar\phi^i)(\phi^j+\bar\phi^j)\,, \cr
{\cal N}_{Ti}&
=i e^\Khat\,(S+\bar S)^2(U+\bar U)(\phi^i+\bar \phi^i)\,, \qquad
{\cal N}_{Ui}
=i e^\Khat\,(S+\bar S)^2(T+\bar T)(\phi^i+\bar \phi^i) \,,\cr
{\cal N}_{SS}&
=-{\textstyle{i\over 2}}{(U+\bar U)(T+\bar T)
	-(\phi^j+ \bar\phi^j)^2\over S+\bar S}\,,\cr
{\cal N}_{ST}&
=-{\textstyle{i\over 2}} (U-\bar U)\,,\qquad
{\cal N}_{SU}
=-{\textstyle{i\over 2}} (T-\bar T)\,,\qquad
{\cal N}_{Si}
= i(\phi^i-\bar\phi^i)\,.\cr
}\label{gaugemoddil}
\eeq
while the couplings of the graviphoton are
\beq
\eqalign{
{\cal N}_{00}&
=-2i\bar S(\bar T\bar U- \bar\phi^{j\,2})\cr
&\hskip 5mm +{\textstyle{i\over 2}} e^\Khat
	\Big[S(TU-\bar T\bar U-\phi^{j\,2}+\bar\phi^{j\,2} )
	- \bar S(\bar TU+T\bar U+2\bar T\bar U-2|\phi^j|^2-2\bar\phi^{j\,2})
	\Big]^2\,,\cr
{\cal N}_{0S}&
=-{\textstyle{1\over 2}}{S(TU+\bar T\bar U-\phi^{j\,2}-\bar\phi^{j\,2})
	-\bar S(T\bar U+\bar TU-2|\phi^j|^2)\over S+\bar S} \,,\cr
{\cal N}_{0T}&
=- \bar S\bar U -{\textstyle{1\over 2}}e^\Khat \, (S+\bar S)(U+\bar U)\cr
&\hskip 2.3cm\times\big[ S (TU-\bar T\bar U- \phi^{j\,2}+\bar\phi^{j\,2})
	-\bar S(\bar TU+T\bar U+2\bar T\bar U- 2|\phi^j|^2-
		2 \bar\phi^{j\,2})\big]\, ,\cr
{\cal N}_{0U}&
=-\bar S\bar T -{\textstyle{1\over 2}}e^\Khat\,(S+\bar S)(T+\bar T)\cr
&\hskip 2.3cm \times\big[S(TU-\bar T\bar U-\phi^{j\,2}+\bar\phi^{j\,2})
	-\bar S(\bar TU+T\bar U+2\bar T\bar U-2|\phi^j|^2-2\bar\phi^{j\,2})
	\big]\, ,\cr
{\cal N}_{0i}&
=2 \bar S\bar\phi^i +e^\Khat\,(S+\bar S) (\phi^i+\bar\phi^i)\cr
&\hskip 2.3cm \times\big[S(TU-\bar T\bar U-\phi^{j\,2}+\bar\phi^{j\,2})
	- \bar S(\bar TU+T\bar U+2\bar T\bar U-2|\phi^j|^2
		-2\bar\phi^{j\,2})\big] \, .
}\label{gaugegp}
\eeq
We observe that all these gauge couplings are indeed non-holomorphic
functions of the moduli, which is a direct consequence of the
mixing between the graviphoton and the Abelian vector superpartners
of the moduli scalars ({\it cf.} the second term in eq.~(\ref{Ndef}).%
\footnote{%
    The non-holomorphicity of the tree-level gauge couplings
    is not present in $N=1$ supersymmetric orbifolds of the
    toroidal compactifications.
    Indeed, in such orbifolds
    the four $U(1)$ gauge bosons   related to $T^2$  disappear
    from  spectrum since the corresponding vertex operators are not
    invariant under the orbifold twist.
    Similarly, all possible Wilson-line moduli $\phi^i$ of the
    compactification are not  twist
    invariant and thus also disappear from the spectrum.
    Therefore, after the $N=1$ truncation, {\em all} the
    gauge couplings are given by simply $S$,
    which is  the well-known property of the tree-level gauge
    coupling in $N=1$ heterotic vacua.}
Furthermore, since $e^{\Khat}\propto(S+\bar S)^{-1}$,
one can easily see that most of the ${\cal N}_{IJ}$ in
eqs.~(\ref{gaugemoddil}) and (\ref{gaugegp})
are proportional to the dilaton's expectation value
and hence the  corresponding gauge couplings become weak in the
large-dilaton limit.
The exceptions are ${\cal N}_{SS}$, which is proportional to
$S+\bar S$ and the off-diagonal matrix elements ${\cal N}_{S\,(T,
U,i\,{\rm or}\,0)}$,
which are of the order $O(1)$ in the large-dilaton limit.
On the other hand, from the string theory we know that {\em all}
the {\em physical} low-energy couplings become weak in the
large-dilaton limit, which suggests that the strongly-coupled
$F_{\m\n}^{+S}$ field
strength in the dilaton $N=2$ superfield should be replaced
with its dual (which is weakly coupled in the large-dilaton limit).
In $N=2$ terms, this is achieved by the symplectic transformation
$(X^I, F_J) \rightarrow (\P^I, \Q_J)$ where
\beq
\eqalign{
\P^I = X^I\quad {\rm for}\  I \neq 1\ , \qquad \P^1 =  F_1, \cr
\Q_I = F_I \quad  {\rm for}\  I \neq 1\ , \qquad \Q_1 = -  X^1\ .}
\label{newbasis}
\eeq
The corresponding symplectic matrix ${\cal O}^\prime$, defined by
\beq
\pmatrix{\hat X{}^I\cr \hat F_J\cr} = {\cal O}^\prime\,
\pmatrix{X^K\cr F_L\cr} \,,
\eeq
thus has nonzero elements
\beq
U^{\prime I}{}_{\!J}=V^\prime_J{}^I=\d^I_J\quad {\rm for} \ I,J\not=
1\,,\quad Z^{\prime 11}=1\,,\quad
W^\prime_{11} =-1\,.
\eeq
The new coordinates $\hat X{}^I$ are, however, not independent,
as they no longer depend on $X^1$. This reflects itself in the
constraint
\beq
\eta_{IJ} \P^I \P^J\ \mathrel{\mathop{=}\limits^{\rm def}}\
\P^1 \P^0 + \P^2 \P^3
- \P^i \P^i - \P^a \P^a\  =\ 0  \
\label{Pconstraint}
\eeq
(the first equality here defines the symmetric matrix $\eta$),
which can be easily verified by an explicit calculation.\footnote{%
    The $\Q_I$ are related to the $\P^I$ according to
    $\Q_I= -2iS\,\eta_{IJ}\P^J$ and themselves satisfy
    a quadratic constraint $\eta^{IJ}\Q_I\Q_J =
    4\Q_0\Q_1+4\Q_2\Q_3-\Q_i\Q_i-\Q_a\Q_a$.
    However, unlike eq.~(\ref{Pconstraint}), which remains valid
    in perturbation theory, the relations involving the
    $\Q_I$ are modified by the loop corrections.
    We shall return to this issue in section~4.2.
    }
Consequently the matrix ${\cal S}^I{}_J(X)=\del\P^I/\del X^J$
has zero determinant and hence no meaningful
prepotential $\hat F(\P)$ can be defined \cite{Ceresole}.
Nevertheless,  the gauge couplings and the K\"ahler potential
for the moduli can be computed in the new
basis from eq.~(\ref{nchange}).
One finds \cite{SFetal,Ceresole}
\beq
\eqalign{
\hat{K}_\mod  &= \Khat= - \log(S + \bar S) - \log 2  (\hat{z}^J
\eta_{JI} \hat{\bar z}^I)\cr
\hat{{\cal N}}_{IJ} &= -2i \bar S\, \eta_{IJ} +2i(S + \bar S)\,
{\eta_{IK}\,\eta_{JL}(\hat{z}{}^K \hat{\bar z}{}^L + \hat{\bar
z}{}^K \hat{z}{}^L)
\over  \hat{z}{}^K \eta_{KL} \hat{\bar z}{}^L} \ ,}
\label{goods}
\eeq
which has a rather symmetric form
in terms of special coordinates $\hat{z}^P \equiv \P^P/\P^0$.
In particular, in the new basis, all the ${\rm Im}~\hat{\cal N}_{IJ}$
are proportional to $S+\bar S$ and hence {\em all} the gauge
couplings become weak in the large-dilaton limit.
Note that the equality $\hat K_\mod = K_\mod$ holds by virtue of
the fact that the
symplectic reparametrization (\ref{newbasis}) does not
involve $X^0$.

The basis $(\hat X{}^I,\hat F_J)$ is particularly well
suited for the treatment of the the target-space-duality symmetries
of generic $N=2$ heterotic string vacua
since the classical Lagrangian is manifestly invariant
under symplectic transformations with $\hat W=\hat Z=0$ and $\hat
U$ (and thus $\hat V$) belonging to $SO(2,2+P)$. This group
follows from the requirement that the tensor $\eta_{IJ}$ is left
invariant: $\hat U{}^{\rm T}\eta\,\hat U= \eta$. In that case
the constraint (\ref{Pconstraint}) is manifestly invariant, while
the K\"ahler metric and the gauge-coupling matrix $\hat {\cal
N}_{IJ}$ transform covariantly. Under this symmetry, the periods
thus transform according to
\beq
\P^I \rightarrow \hat U{}^I{}_J \,\P^J\ , \qquad \Q_I \rightarrow
\big[\hat U{}^{-1}\big]^J{}_I\, \Q_J\,,
\label{Linv}
\eeq
while the field strengths and vector potentials
also transform according to the $\hat U$ matrix.
The dilaton field remains invariant at the classical level (see,
however, the discussion in section~4.2).

Beyond the tree level, the continuous $SO(2,2+P)$ symmetry group of the
low-energy effective theory is explicitly broken by the string loop
corrections, but its maximal discrete subgroup $SO(2,2+P,{\bf Z})$ (or a
subgroup thereof)
remains an exact symmetry of the underlying string vacuum family
\cite{SEN}
--- the target-space duality of the Narain lattice ---
and hence should be manifest in the low-energy EQFT as well.
In the following sections we shall discuss the constraints imposed by
this discrete symmetry upon the loop corrections to the holomorphic
prepotential $\Fc$.
For the moment, let us simply make a few comments regarding the
target-space duality group for toroidal compactifications. In
that case this group contains $PSL(2,{\bf Z})_T\times
PSL(2,{\bf Z})_U$ and the action of
the target-space duality group on the moduli can
be determined from the heterotic string compactification as
described in section 2.
We recall that the first factor of the discrete group $PSL(2,{\bf
Z})_T\times PSL(2,{\bf Z})_U$ of  toroidal compactifications,
acts on the moduli as
\begin{equation}
T\rightarrow {aT-ib\over icT+d}\,,\qquad
U\rightarrow U-{ic\over icT+d}{\phi^i\phi^i}\ , \qquad
\phi^i\rightarrow {\phi^i\over icT+d}\ ,
\label{Tduality}
\end{equation}
while $S$ remains invariant. The corresponding symplectic
matrices (in the  basis $(\hat X^I,\hat F_I)$), are given by
\begin{equation}
\hat U= \pmatrix{d&0&c&0&0 \cr 0&a&0&-b&0 \cr b&0&a&0&0 \cr 0&-c&0&d&0\cr
0&0&0&0&{\bf 1}_P}\,,
\quad
\hat V= (\hat U{}^T)^{-1} = \pmatrix {a&0&-b&0&0 \cr 0&d&0&c&0
\cr -c&0&d&0&0 \cr 0&b&0&a&0\cr 0&0&0&0&{\bf 1}_P}\,,
\label{dualmat}
\end{equation}
while $\hat W =\hat Z =0$.
These matrices can be obtained in many ways, but by far the easiest
procedure is to straightforwardly embed the $SL(2,{\bf Z})_T$ into
the $SO(2,2)$ subgroup of the $SO(2,2+P)$ group of classical
$\hat U$ matrices. The two facts to remember is that the $\P^I$
should transform linearly into each other and that the field $U$ should
be inert in the absence of $\phi^i$.
Alternatively one may employ the nilpotent subgroups
constructed in subsection~3.1.

Although the discussion thus far was confined to the moduli, at
the classical level there is no
essential difference between  the moduli and the charged scalars
$\mat^a$, as one can see directly from the $F$-function in (\ref{ourF}).
Therefore, the previous symmetry consideration can be easily extended
to include  the $X^a$  which are inert under  $SO(2, P+2)$.
Consequently, the $\mat^a$ transform according to
\beq
\mat^a \rightarrow {\mat^a\over icT+d}
\label{mattertrans}
\eeq
under the $PSL(2,{\bf Z})_T$ transformations.

Beside the target-space duality transformations, the classical field
equations of the effective low-energy theory (but not its Lagrangian)
are invariant under the so-called $S$-duality transformations
\cite{SDUAL,SEN}.
These transformations form an $SL(2,{\bf Z})$ group, which
act on the dilaton field $S$ according to
\begin{equation}
S\rightarrow {aS- ib\over icS+d}\,,
\quad a,b,c,d\in{\bf Z},\quad ad-bc=1
\label{Sdual}
\end{equation}
while the other scalar fields $z^2,\ldots,z^n$ remain invariant.
The corresponding $Sp(8+2P,{\bf Z})$ matrices in the basis
$(\P^I,\Q_I)$ are
\begin{equation}
\hat U{}^I_{\,J}= d\  \d^I_{\,J} \,, \qquad \hat V_J{}^I= a\
\d^I_{\,J}\,, \qquad
\hat W_{IJ} = -2 b \  \eta_{IJ} \,, \qquad \hat Z{}^{IJ}= -
{\textstyle{1\over 2}}c\ \eta^{IJ} .
 \label{sss}
\end{equation}
This result follows straightforwardly from the tree-level relation
$\Q_I=-2iS\eta_{IJ}\P^J$, which demonstrates yet another advantage
of the $(\P^I,\Q_I)$ basis.
Of course, the same result can be obtained in the original
$(X^I,F_I)$  basis as well in a variety of ways,
for example, one may use two subgroups of the $SL(2,{\bf Z})_S$,
one with $a=d=1$, $c=0$, the other with $a=d=1$, $b=0$, which
together generate the entire $SL(2,{\bf Z})_S$,
for which the respective transformations are precisely those
corresponding to
(\ref{btransform}) and (\ref{atransform}). As
the symplectic matrices (\ref{btransform}) and (\ref{atransform})
are in the $(X^I,F_J)$ basis, we must use ${\cal O}={\cal
O}^\prime\hat{\cal O}\,{\cal O}^{\prime -1}$ in order to reproduce
eq.~(\ref{sss}).
Needless to say, however, is that all of the above results, for
both the $S$-duality and the target-space dualities, can be
independently derived from string theory \cite{FLT,Shevitz,FKLZ}.

Among the $S$-dualities (\ref{Sdual}),
of particular interest is the shift $S\to S-i$, which affects
no physical couplings except the $\theta$ angles
which are shifted by $2\pi$; this is the {\em discrete} Peccei-Quinn
symmetry, which we  assume to be exact at the quantum level.\footnote{%
    We apologize for a rather cavalier normalization of the gauge
    couplings.
    For the dilaton field $S$ normalized in accordance with
    eq.~(\ref{Sdual}) for the $S$-dualities, the conventionally
    normalized tree-level gauge couplings should be $g_\ind^{-2}=
    4\pi k_\ind\,Re\,S$, where $k_\ind$ is the level of the
    Ka\v{c}-Moody algebra giving rise to the gauge group $G_\ind$.
    }
On the other hand, the  transformation
$S\rightarrow 1/S$ interchanges all the electric and magnetic
$U(1)$ field strengths and inverts all the gauge couplings
(cf.~eq.(\ref{Ssymp})).
Correspondingly, the electric and magnetic charges are also
interchanged, which means that this duality
mixes elementary string states with  the non-perturbative
solitons and  therefore is of inherently non-perturbative nature.
It is presently unknown which of the $S$-duality transformations
are true symmetries of the  quantum theory, but it is clear
that quantum corrections to the holomorphic prepotential
necessarily modify the explicit form (\ref{Sdual}) of
such transformations. We expect the corresponding symplectic
matrices to constitute a subgroup of the matrices
(\ref{sss}).

\subsect{Perturbative corrections}

This section is about the perturbative corrections to the prepotential
for heterotic string vacua.
As we argued in section~3.2, at the quantum level
the distinction between moduli and non-moduli scalars
becomes important due to their very different
renormalization behavior.
Therefore, it proves convenient to expand  $\Fc$ around
small $\mat^a$  as in eq.~(\ref{Fexp});
at the tree level the moduli-dependent
coefficients of this expansion are determined in eq.~(\ref{treecouplings})
and one is  left with
\beq
\eqalign{
\fone &= -S(TU - \sum_i \phi^i \phi^i) + \foneol (T,U,\phi^i)\, , \cr
\ftwo_\ind  &= S +  \ftwool_\ind (T,U,\phi^i)\ . }
\label{hfexp}
\eeq
These formul\ae, in which both $\foneol$ and $\ftwool$ are functions
of all the moduli {\em except} $S$,
uses the fact that the dilaton serves
as the loop-counting parameter of the heterotic string.
For the same reason, any possible two-loop or higher-loop
corrections would have to be proportional to negative powers
of the dilaton and because of the continous Peccei-Quinn symmetry
(which persists to all orders in the perturbation theory),
such corrections would have to involve the negative powers of
the $(S+\bar S)$ combination rather than just $S$.
On the other hand, $\bar S$ clearly cannot appear in the
holomorphic prepotential $\Fc(\mod)$ and hence in string theory,
all perturbative corrections to the prepotential stop at the
one-loop level, in full analogy to the field-theoretical
expansion (\ref{Floop}), which also terminates at
the one-loop order.
In the $N=1$ context, the same argument forbids two- or higher-loop
corrections to the Wilsonian gauge couplings \cite{nilles};
for $N=2$ this non-renormalization theorem is more powerful
since the prepotential $\Fc$ determines both the Wilsonian gauge
couplings and the K\"ahler potential.
Furthermore, by similar arguments it follows that the couplings
of the hypermultiplets are not corrected at all at any loop order
and hence the tree-level hyper-moduli space is the exact
hyper-moduli space to all orders in perturbation theory.

Neither $\foneol (T,U,\phi^i)$ nor $\ftwool_\ind (T,U,\phi^i)$
can be  arbitrary functions of the moduli, since
they should respect
any exact duality  symmetry a string vacuum might have.
In the previous section we saw that the tree-level geometry of the
moduli space is invariant under the $SO(2,2+P)$ isometry group,
and from string theory we know that transformations belonging to
a discrete $SO(2,2+P,{\bf Z})$ subgroup of this isometry group
are in fact exact symmetries of string vacua to all orders
in perturbation theory.
The goal of this and the following sections is to find the precise
conditions such exact symmetries impose on the holomorphic functions
$\foneol$ and $\ftwool_\ind$.
For the present section, we assume a completely generic $N=2$ vacuum
family of the heterotic string and keep our discussion as general
as possible.
In the following section
we then specialize to toroidal compactification.

At the quantum level of the effective field theory,
the Wilsonian Lagrangian does  not necessarily
share the quantum symmetries; only
the physical, effective couplings have
to be invariant functions of the moduli.
Let us therefore begin with the moduli multiplets, for which
the Wilsonian gauge couplings are equal to the effective couplings
and hence are invariant or rather covariant under the exact
modular symmetries of the string theory.
Suppressing the non-moduli vector multiplets from our notations,
we write the holomorphic prepotential for the remaining homogeneous
variables $X^I$ ($I=0,1,\ldots,P+3$) as
\beq
F(X)= H^{(0)}(X) + \Hone(X)\,,
\label{Fdecom}
\eeq
where $H^{(0)}(X)$ is the tree-level prepotential (\ref{ourF})
while  $\Hone(X) = -i(X^0)^2\,\foneol$ represents the one-loop
contribution.
Both functions are homogeneous of
second degree and according to (\ref{hfexp}) $\Hone$ does not
depend on  $X^1$.
However, the most convenient variables for our  purpose are again
$\P^I$ and $\Q_I$.
Since the one-loop prepotential $\Hone$ does not depend on $X^1$,
it follows that $\P^1=F_1$ is not modified by loop
corrections.
Hence, in the quantum theory the $\P^I$ satisfy exactly the same
constraint~(\ref{Pconstraint}) as in the classical case.
On the other hand, the relation between the $\Q_I$ and the $\P^I$
is sensitive to the one-loop prepotential $\Hone$ and we now have
\beq
\Q_I\ =\ -2i S\, \eta_{IJ} \P^J\ +\ \Hone_I\,,
\label{Folcorr}
\eeq
where $\Hone_I \equiv \del\Hone(X)/\del X^I$
(cf.~eqs.~(\ref{newbasis}) and (\ref{Fdecom})).
Obviously $\Hone_1=0$, so that all the
$\Q_{I\neq1}$ are modified by the quantum corrections, but
$\Q_1$ keeps its classical value $\Q_1=-X^1=-iS\P^0$.
For the same reason, $\Hone_I\P^I=\Hone_I X^I$;
at the same time, $\Hone_I X^I=2\Hone$ because of the homogeneity of the
function $\Hone(X)$.
Combining these two facts with eq.~(\ref{Folcorr}) and with the
constraint~(\ref{Pconstraint}), we arrive at
\beq
\Hone(X)\ =\ \textstyle{1\over2} \Q_I\P^I ,
\label{HdefPQ}
\eeq
which expresses the one-loop prepotential directly in terms of the
symplectic variables $(\P^I,\Q_I)$.

In perturbative string theory, the moduli fields
$T$, $U$ and $\phi^i$ have fixed relations to their vertex operators
and hence the transformation rules for these fields are
completely determined at the tree level and are not corrected
by the string loops.
In terms of the $(\P^I,\Q_I)$ variables of the EQFT, this means
that the $\P^I$ should transform exactly as in the classical
theory (cf.~\ref{Linv}),
without any perturbative corrections.
On the other hand, the corresponding transformation rules
(\ref{Linv}) for the $\Q_I$ become
modified at the one-loop level since the Lagrangian is no longer
invariant. Instead the transformation rules have to generate discrete
shifts in various $\theta$ angles due to monodromies around semi-classical
singularities in the moduli space where massive string modes become
massless.
We have anticipated this situation in eqs.~(\ref{Tsymp}):
Instead of the classical transformation rules (\ref{Linv}),
in the quantum theory, $(\P^I,\Q_I)$ transform according to
\beq
\P^I\ \rightarrow\   \hat U{}^I_{\,J}\, \P^J,\qquad
\Q_I\ \rightarrow\   \hat V_I{}^J \,\Q_J\ +\  \hat W_{IJ}\, \P^J\ ,
\label{oltrans}
\eeq
where
\beq
\hat V = (\hat U^{\rm T})^{-1},\quad
\hat W = \hat V \Lambda\,,\quad \Lambda=\Lambda^{\rm T}
\eeq
and $\hat U$ belongs to  $SO(2,2+P,{\bf Z})$.
Classically, $\Lambda=0$, but in the quantum theory, $\Lambda$
is an arbitrary real symmetric matrix, which should be integer
valued in some basis
(but not necessarily in the basis of $\P^I$ defined in eqs.\
(\ref{newbasis}) and (\ref{modid})) so that the ambiguities in
the $\theta$ angles  are discrete
($\delta\theta=\hat W\hat U^{-1}=\hat V\Lambda\hat V^{\rm T}$).
In particular, for a closed monodromy around a singularity,
$\P^I\to\P^I$ {\it i.~e.}\ $\hat U=1$,
but $\Lambda\neq0$ and $\Q_I\to\Q_I+\Lambda_{IJ}\P^J$.

We recall that the prepotential itself is in general not invariant
under a symmetry of the equations of motion corresponding to the
effective action, as one can easily verify for the
tree-level results of the previous section, but the period
transformation rules are correctly induced by the transformations
of the coordinates. Therefore substituting the period
transformations (\ref{oltrans}) into eq.~(\ref{HdefPQ}), one
immediately obtains the corresponding transformation rule for
$\Hone$,
\beq
\Hone(\tilde X)\ = \Hone(X)\ +\  \textstyle{1\over2} \Lambda_{IJ} \P^I \P^J .
\label{Hshift}
\eeq
Note that the dilaton field does not appear anywhere in this formula.
To put the symmetry relation (\ref{Hshift}) in its proper context,
it is important to keep in mind that $\Hone$ should have
a logarithmic singularity whenever an otherwise massive string
mode becomes massless;
therefore, as an analytic function of $(X^0,X^2,\ldots,X^{P+3})$,
$\Hone(X)$ is generally multi-valued.
According to eq.~(\ref{Hshift}), the ambiguities of $\Hone$ amount
to quadratic polynomials in the variables $\P^I$ with some discrete
real coefficients\footnote{%
    In terms of $\foneol(T,U,\phi)$, the ambiguities are quadratic
    polynomials in variables $1$, $iT$, $iU$, $i\phi^i$ and
    $(TU-\sum\phi^2)$ with discrete {\sl imaginary} coefficients.
    Note that in terms of $T$, $U$ and $\phi^i$, such polynomials
    are quartic.%
    };
indeed, under a closed monodromy one generally has
$\Hone\to\Hone+{1\over2}\Lambda_{IJ}\P^I\P^J$ even though the fields
$(X^0,X^2,\ldots,X^{P+3})$ remain unchanged.
However, modulo these ambiguities, $\Hone$ should be invariant
under all the exact symmetries of the perturbative string theory.
This is the main result of this section.

Let us now turn our attention to the dilaton field $S$.
In perturbative string theory, the dilaton vertex and its superpartners
have fixed relations to the vector-tensor multiplet.
However, the duality relation between this vector-tensor multiplet and
the Abelian vector multiplet containing $S=-iX^1/X^0$ is not fixed but
suffers from perturbative corrections in both string theory and field theory.
Therefore, while the vector-tensor multiplet is inert under all the
perturbative symmetries of the string's vacuum,
the $S$ field is only invariant classically but has non-trivial
transformation properties at the one-loop level of the quantum theory.
Indeed, using the relation $X^1=-\Q_1=-iS\P^0$, it is easy to show that
the transformation rules (\ref{oltrans}) imply
\beq
S\ \rightarrow\ \tilde S \ =\  S\
+\ {i\hat V_1{}^J\left(\Hone_J +\Lambda_{JK}\P^K\right)
	\over \hat U^0{}_I\,\P^I }\,,
\label{diltrans}
\eeq
which in turn is sufficient to assure the correct transformation
properties of all the $\Q_I$ and not just the $\Q_1$.\footnote{%
    Using a convenient identity $(\del\P^1/\del X^I)=\delta^1_I-
    2\eta_{IJ}(\P^J/\P^0)$, one can show that eq.~(\ref{Hshift})
    implies the following transformation
    rule for the first derivatives of $\Hone$:
    \beq
    \tilde \Hone_I\ =\ \del\tilde\Hone/\del\tilde X^I\
    =\ \hat V_I{}^J\,H_J + \hat W_{IJ}\,\hat X{}^J
    +2i(\tilde  S-S) \eta_{IJ} \hat U{}^J_{\,K}\,\hat X{}^K \,,
    \nonumber\end{equation}
    where the difference $(\tilde S-S)$ is precisely as in
    eq.~(\ref{diltrans}) (note that this difference does not depend on the
    dilaton itself but only on the other moduli).
    It is easy to see that this transformation rule is precisely
    what is needed to assure complete consistency between eqs.\
    (\ref{Folcorr}) for the $\Q_I$ and the transformation rules
    (\ref{oltrans}).
    }
However, if one does not insist upon the dilaton field being a
{\em special} coordinate of the $N=2$ supersymmetry, it is possible
to define a modular-invariant dilaton-like complex field by simply
shifting $S$ by a function of the other moduli.
Specifically,\footnote{%
    As usual, $\Hone_{IJ}$ denote the second derivatives
    $\del_I\del_J\Hone$.
    Note that since $\Hone_1=0$, the contraction $\eta^{IJ}\Hone_{IJ}$
    involves only $I,J=2,\ldots,P+3$.
    The transformation rules for the $\Hone_{IJ}$ can be derived in
    the same manner as the rules for the first derivatives $\Hone_I$
    in the previous footnote (although the algebra is somewhat more
    complicated).
    }
\beq
\hatS\ =\ S + {1\over 2(P+4)}\left[
	i\eta^{IJ}\,\Hone_{IJ} \ +\ \Ll \right] ,
\label{dilinv}
\eeq
where $\Ll$ is a holomorphic function of the moduli
whose duality transformation rules amount to imaginary constant shifts
\beq
\Ll\ \rightarrow\  \Ll\ -\  i \eta^{IJ} \Lambda_{IJ}\ .
\label{Lshift}
\eeq
In the following subsection we shall see that such a function is
necessary to keep $\hatS$ finite.

In $N=1$ supersymmetric vacua of the heterotic string, $S$
belongs to a chiral supermultiplet dual to a linear multiplet.
The linear multiplet has a fixed relation to the string vertices
and is therefore inert under all the perturbative symmetries,
while the $S$ field has to be constructed order by order in
perturbation theory.
However, in the $N=1$ case one is free to redefine $S$ by adding
to it an arbitrary holomorphic function of the other moduli.
This ambiguity is inherent in the chiral multiplet--linear multiplet
duality relation and one may use it to define a modular invariant
chiral superfield for the dilaton according to some analogue
of eq.~(\ref{dilinv}).
By contrast, the $N=2$ supersymmetry does not allow for non-linear
redefinition of the vector supermultiplets.
Therefore, while the modular-variant scalar $S$ is a legitimate
member of an $N=2$ vector multiplet, the modular-invariant $\hatS$
is not and hence cannot be simply used in place of the $S$.

Finally, consider the non-moduli gauge couplings $\ftwo_\ind(\mod)$.
Clearly, the physical gauge couplings $g^{-2}_\ind(p^2)$ have to
respect all the exact symmetries a string vacuum might possess.
The Wilsonian gauge couplings, however, are not always invariant
because they act as local counterterms compensating for potential
anomalies of some of the symmetries \cite{KLa,HOLAN,KLb}.
For the local $N=2$ supersymmetry, such anomalies are associated with
non-trivial transformations
of the K\"ahler function $\Khat$ in eq.~(\ref{anofin}).
Indeed, under a generic $SO(2,P+2,{\bf Z})$ symmetry (\ref{oltrans}),
the K\"ahler function~(\ref{KP}) transforms according to
\beq
\eqalign{
\Khat(\mod,\modb)\ \rightarrow\ {} &
\Khat(\mod,\modb)\ +\ \log|\tilde{X}^0/X^0|^2\cr
{}=\ {}&
\Khat(\mod,\modb)\ +\ \log|\hat U^0{}_J\P^J/\P^0|^2 .\cr }
\eeq
Hence, in order to keep the physical coupling $g_\ind^{-2}$
in eq.~(\ref{anofin}) invariant,
the Wilsonian coupling $\ftwo_\ind$ should transform as
\beq
f_\ind(\mod)\ \rightarrow\ f_\ind(\mod)\
-\ {b_\ind \over 8 \pi^2} \,\log(\hat U{}^0{}_J \P^J/\P^0)\,.
\label{fcomp}
\eeq
This one-loop modification of the modular transformation
properties of the $\ftwo_\ind$ is entirely analogous to
the situation encountered in $N=1$ supersymmetry.
However, while in the $N=1$ case one may keep the dilaton $S$ invariant
and attribute the entire anomaly to the one-loop gauge couplings
$\ftwool_\ind(\mod)$,
in the present $N=2$ case one has to live with the dilaton $S$
transforming according to eq.~(\ref{diltrans}).
Consequently, the transformation rule for the $\ftwool_\ind(\mod)$ is given
by the difference between eqs.~(\ref{fcomp}) and (\ref{diltrans}).

\subsect{Toroidal Compactifications}

Thus far our analysis was generic and applicable to any
$N=2$ vacuum of the heterotic string.
Let us now apply this general formalism to the concrete case
of  toroidal compactifications of six-dimensional
$N=1$ string vacua.
We begin by turning off all the Wilson-line parameters $\phi^i$;
more precisely, we consider the domain of small $\langle\phi^i\rangle$
in which $\phi^i$ act as matter fields $\mat^a$ rather than moduli,
and our goal is to compute the quantum corrections $\foneol$ and
$\ftwool_\ind$ as functions of the toroidal moduli $T$ and $U$.
For the case at hand, the target-space duality group is
$SO(2,2,{\bf Z})$ consisting of the $T\leftrightarrow U$ exchange and
of the $PSL(2,{\bf Z})_T$ and $PSL(2,{\bf Z})_U$ dualities whose action
is described by eqs.~(\ref{Tduality}) and~(\ref{dualmat}).
Substituting these dualities (for $\phi^i=0$) into
the general transformation laws
(\ref{Hshift})--(\ref{fcomp}) of  the previous section,  we find
\beq
\eqalign{
T\ & \to\ {aT-ib\over icT+d}\,, \qquad U\ \rightarrow\ U , \cr
\foneol (T,U)\ & \to\ {\foneol (T,U)+\shift(T,U)\over (icT+d)^2}\,, \cr
\ftwo_\ind(S,T,U)\ & \to\ \ftwo_\ind(S,T,U)\
	-\ {b_\ind \over 8\pi^2}\, \log(icT+d) \cr
}
\label{Tonh}
\eeq
and a similar set of transformations
(with $T$ and $U$ interchanged) for the $PSL(2,{\bf Z})_U$.
The appearance of
$\shift={i\over2}\Lambda_{IJ}\P^I\P^J/(\hat X^0)^2$
in these formul\ae\ complicates the symmetry properties of
the one-loop moduli prepotential, which would otherwise be a modular function
of weight $-2$ with respect to both $T$ and $U$ dualities.
However, $\shift$ is a  quadratic polynomial  in the variables
$(1,iT,iU,TU)$ and hence $\del_T^3\shift=\del_U^3\shift=0$;
also, it is a mathematical fact that the third derivative of a
modular function of weight $-2$ is itself a modular function of
weight $+4$ even though the derivative is ordinary rather than covariant.
{}From these two observations, we immediately learn that
$\del_T^3\foneol(T,U)$ is a single-valued modular function
of weight $+4$ under the $T$-duality and of weight $-2$ under
the $U$-duality and there are no anomalies in its modular transformation
properties;
the same is of course true for the $\del_U^3\foneol$, with the two
modular weights interchanged.

The exact analytic form of a modular function can often be completely
determined from the knowledge of its singularities and its
asymptotic behavior when $T\to\infty$ or $U\to\infty$.
It was argued in ref.~\cite{KLb} that the gauge couplings of an $N=1$
orbifold cannot grow faster than a power of $T$ or $U$ in any
decompactification limit and the same argument applies here to the
one-loop prepotential $\foneol$ and any of its derivatives.
Let us therefore consider the singularity structure of the $\foneol(T,U)$.

The gauge couplings of the $[U(1)_L]^2$ containing
the vector partners of $T$ and $U$ become singular whenever there
are additional massless particles charged under this group.
As discussed in section~2, this happens along the complex lines
$T\equiv U$, where the $U(1)^2_L$ group
is enlarged to an $SU(2)\times U(1)$\footnote%
    {As in section~2, by $T\equiv U$ we mean that $T$ and $U$ are
    equal modulo an $SL(2,{\bf Z})$ transformation.%
    };
when such lines intersect each other, the group is further enlarged
to an $SU(2)\times SU(2)$ (at $T\equiv U\equiv 1$) or an $SU(3)$
(at $T\equiv U\equiv\rho=e^{2\pi i/12}$).
However, for a fixed generic value of $U$, the only singularities
in the complex $T$-plane (or rather half-plane ${\rm Re}\, T>0$) are at
$T\equiv U$ while the points $T\equiv 1\not\equiv U$ and
$T\equiv\rho\not\equiv U$ are perfectly regular;
the same is of course true for the singularities in the $U$-plane
(or rather half-plane) when $T$ is held fixed at a generic value.
Furthermore, the singular part of the prepotential along the $T=U$
line is completely determined by eq.~(\ref{FONEphi}), in which
we should identify $\phi={1\over2}(T-U)$, $\mod={1\over2}(T+U)$
and $b_{(\phi)}=-4$ (for an $SU(2)$ without any non-singlet
hypermultiplets).
Hence, for generic $T$ or $U$ but small $T-U$,
\beq
\fone(T\approx U)\ =\ \textstyle{1\over16\pi^2}\,
(T-U)^2\log(T-U)^2\ +\ {\rm regular},
\label{fonesing}
\eeq
although the ``regular'' term here is only regular when
$T\approx U\not\equiv 1,\rho$.
Note that $\fone$ is singular but finite when $T\approx U$; its
third derivatives $\del_T^3\fone=\del_T^3\foneol$
and $\del_U^3\fone=\del_U^3\foneol$ have simple
poles at that point and similar poles whenever
$T\equiv U\!\!\pmod{SL(2,{\bf Z})}$.
This fact, plus all the other properties of the functions
$\del^3_{T,U}\foneol(T,U)$ we have stated above, allow us to
uniquely determine%
\footnote{%
    In eq.~(\ref{hfinal}), $\eta$ is Dedekind's eta-function,
    $E_4$ and $E_6$ are the normalized Eisenstein's modular
    forms of respective weights $+4$ and $+6$ and $j$ is
    the modular invariant function $j=E_4^3/\eta^{24}$.
    The arguments of these functions are $iT$ and $iU$ because
    mathematicians' conventions differ from the
    string-theoretical conventions used in this article.%
    }
\beq
\eqalign{
\del_T^3\foneol\ &
=\ {+1\over2\pi}\,{E_4(iT)\, E_4(iU) E_6(iU) \eta^{-24}(iU)
	\over j(iT)\,-\,j(iU)}\,,\cr
\del_U^3\foneol\ &
=\ {-1\over2\pi}\,{E_4(iT) E_6(iT) \eta^{-24}(iT)\, E_4(iU)
	\over j(iT)\,-\,j(iU)}\,.\cr
}\label{hfinal}
\eeq
This formula obviously determines the function $\foneol(T,U)$
itself up to a polynomial $\shift$ that is at most quadratic
in $T$ and in $U$, but we are unfortunately unable to write
that function in terms of familiar modular functions.
However, it is easy to see that eqs.~(\ref{hfinal}) imply
\beq
\del_T\del_U\foneol\
=\ \textstyle{-1\over4\pi^2}\,\log\left( j(iT) - j(iU)\right)\
+\ {\rm finite},
\label{logDJ}
\eeq
which has a curious property that the coefficient of the
logarithmic divergence
is $2/8\pi^2$ when $T\equiv U\not\equiv 1,\rho$
but becomes $4/8\pi^2$ when $T\equiv U\equiv 1$
and $6/8\pi^2$ when $T\equiv U\equiv\rho$,%
\footnote{%
    The derivative of the $j(iT)$ function has a zero when
    $T\equiv1$ and a double zero when $T\equiv\rho$.%
    }
in precise agreement with the number of the massive
string modes that become massless in each case
(respectively, 2, 4 and 6 vector multiplets).
Indeed, in ref.~\cite{CLM1}, this selfsame property of the
heterotic string's vacua was used to determine the singularity
structure of the gauge couplings such as~(\ref{logDJ}).
In this article, however, we arrived at eqs.~(\ref{hfinal}) and
(\ref{logDJ}) by considering only the $T\equiv U\not\equiv 1,\rho$
vacua and the special properties of the $T\equiv U\equiv 1,\rho$
vacua emerged  courtesy of mathematical properties of the modular
functions.
Nevertheless, it is nice to have our result confirmed by an
unrelated string-theoretical argument.

Now consider the dilaton.
As we discussed in the previous section, the special $N=2$
coordinate $S$ for the dilaton field of a quantum theory
is not modular invariant, but there is a non-special coordinate
$\hatS$ that is modular invariant.
The difference between the two coordinates
\beq
\hath(T,U)\ \mathrel{\mathop{=}\limits^{\rm def}}\ \hatS\ -\ S\
=\ -\textstyle{1\over2}\del_T\del_U\foneol(T,U)\
+\ \textstyle{1\over8}\Ll(T,U)
\label{hathdef}
\eeq
({\it cf.}\ eq.\ (\ref{dilinv}))
must be finite throughout the $(T,U)$ moduli space since otherwise
one would not be able to use the value of $\hatS$ as a universal
string-loop counting parameter.
For the same reason, $\hath(T,U)$ should not grow faster than $T$ or $U$
in the decompactification limits $T\to\infty$ and $U\to\infty$.
Combining these restrictions with eq.~(\ref{logDJ}) for the
$\del_T\del_U\foneol$ and with the requirement
(\ref{Lshift}) that $\Ll(T,U)$ should be modular invariant up to
a constant imaginary shift, we immediately arrive at
\beq
\Ll(T,U)\ =\ -\textstyle{1\over\pi^2}\,
   \log \left( j(iT) - j(iU)\right)\ +\ {\rm const},
\label{Lldef}
\eeq
which indeed shifts by an imaginary constant when $T$ (or $U$) circles
a singular line $T\equiv U$.
Notice that although eqs.~(\ref{logDJ}) and (\ref{Lldef})
provide for the {\sl finiteness} of the difference (\ref{hathdef}),
it is nevertheless singular and multivalued
and its derivatives $\del_{T,U}\hath(T,U)$ diverge logarithmically
when $T\equiv U$.
The multi-valuedness of the difference~(\ref{hathdef}) is particularly
disturbing since it implies that $S$ is not only subject to non-trivial
modular transformations but is not even single-valued for given values
of $T$ and $U$.%
\footnote{%
    Note that according to the arguments of the previous section, the
    $\hatS$ coordinate is invariant under all semi-classical symmetries
    of the perturbative string theory,
    including the monodromies that leave $T$ and $U$ invariant;
    in other words, $\hatS$ is both single-valued and modular-invariant.
    On the other hand, the $N=2$ superfield $S$ is multi-valued,
    but any possible ambiguity in $S$ has to be a linear combination of
    the four $i\P^I/\P^0=(i,T,U,iTU)$;
    these are the only ambiguities allowed for the $N=2$ superfields.%
    }
While one should expect such multi-valuedness of $S$ in strongly
coupled non-perturbative gauge theories, it is rather surprising
to discover it already at the one-loop level.

Next consider the Wilsonian gauge couplings $\ftwo_\ind$
for the gauge groups that the toroidal compactification
inherits from the six-dimensional theory, $i.\,e.$,
for all the gauge groups other than $U(1)^4_{L+R}$.
The modular transformation rule for these couplings is given by
the last eq.~(\ref{Tonh}),
which has exactly the same form as its analogues for the
$N=1$ factorizable orbifolds considered in ref.~\cite{KLb}.
Consequently, for exactly the same reasons as in the
$N=1$ case, we now have
\beq
\ftwo_\ind(S,T,U)\ =\ \hatS\
-\ {b_\ind\over 8\pi^2}\left(
	\log \eta^2 (iT) + \log \eta^2 (iU) \right)\
+\ {\rm const}.
\eeq
Note, however, that this formula involves the modular-invariant
coordinate $\hatS$ for the dilaton rather than the $N=2$
special coordinate $S$ that appears in the tree-level term in
eq.~(\ref{hfexp}).
Therefore, in terms of the $N=2$ supermultiplets, the
one-loop corrections to the gauge couplings are
\beq
\ftwool_\ind(T,U)\
=\ -{b_\ind\over 8\pi^2}\left(
	\log \eta^2 (iT) + \log \eta^2 (iU) \right)\
+\ \hath(T,U) .
\label{ffdet}
\eeq
The first term on the right hand side here, plus the K\"ahler
correction according to eq.~(\ref{anofin}), together
constitute precisely the
non-universal string-threshold correction to the gauge couplings
obtained via an explicit string-loop calculation in ref.~\cite{DKL}.
The second term on the right hand side of eq.~(\ref{ffdet})
amounts to a universal threshold correction.
Such universal corrections were disregarded in ref.~\cite{DKL},
but they are also obtainable from string-loop calculations;
we shall return to this point momentarily.

Before that, however, let us consider the loop-corrected K\"ahler
potential $\Khat(S,T,U)$.
Substituting the prepotential (\ref{hfexp}) into eq.~(\ref{Zexp}),
we immediately obtain
\beq
\Khat(S,T,U)\ =\ -\log\left[ S\,+\,\bar S\, +\,V_{GS}(T,U)\right]\
-\ \log(T+\bar T)\ -\ \log(U+\bar U) ,
\label{Kloop}
\eeq
where
\beq
V_{GS}(T,U)\ =\ {2(\foneol+\bar\foneol)\,
-\,(T+\bar T)(\del_T\foneol+\del_{\bar T}\bar\foneol)\,
-\,(U+\bar U)(\del_U\foneol+\del_{\bar U}\bar\foneol)
\over (T+\bar T)\,(U+\bar U)}
\label{GSfunction}
\eeq
is the Green-Schwarz term \cite{HOLAN} describing the mixing
of the dilaton with the moduli $T$ and $U$.
In $N=1$ vacua of the heterotic string such mixing arises at all
loop levels of the string theory (except the tree level, of course),
but in the $N=2$ case it is completely determined at the one-loop
level.
The importance of the Green-Schwarz term has to do with the fact
that in the vector supermultiplet formalism for the dilaton, the
true loop-counting parameter of the heterotic string is neither
$S+\bar S$, nor even $\hatS+\bar\hatS$, but rather
\beq
S\ +\ \bar S\ +\ V_{GS}(T,U)\
=\ \hatS\ +\ \bar\hatS\ +\ \hatV_{GS}(T,U) ,
\label{LoopCounter}
\eeq
($\hatV_{GS}$ is defined by this equation),
which is directly related to the scalar component
of the vector-tensor multiplet (or linear multiplet in the $N=1$ case).
Therefore, a direct one-string-loop calculation of the threshold
corrections $\Delta_\ind$ to the gauge couplings should be
interpreted according to \cite{KLb}
\beq
\left[ g^{-2}_\ind(p^2)\right]^{\scriptscriptstyle{\rm one-loop}}\
=\ {\rm Re}\,\hatS\ +\ {\textstyle{1\over2}} \hatV_{GS}(T,U)\
+\ \Delta_\ind(T,U)\ +\ {b_\ind\over 16\pi^2}\,
\log{M^2_{\rm string}\over p^2}\,.
\eeq
Hence, a direct string calculation of the universal part of all the
$\Delta_\ind$ would immediately yield the modular-invariant Green-Schwarz
term $\hatV_{GS}$, or rather
\beq
\!\eqalign{
\Delta^{\rm univ}\, &
=\, -\textstyle{1\over2}\hatV_{GS}(T,U)\cr
&=\, -{1\over4}
    \left[{2\over T+\bar T}\,-\,(\del_T+\del_{\bar T})\right]\,
    \left[{2\over U+\bar U}\,-\,(\del_U+\del_{\bar U})\right]\,
    \left( \foneol+\bar\foneol\right)\
    +\ {1\over16}\left(\Ll+\bar\Ll\right).\cr
}\!\label{MIGSt}
\eeq
The calculation itself will be presented in a forthcoming article by
some of the present authors; the techniques we have used are rather
similar to those of ref.~\cite{DKL,AGN}.
For the purposes of the present article, let us simply state that
the result is a complicated integral, which can be shown to satisfy
the modular-invariant differential equation
\beq
\eqalign{
\left[ (T+\bar T)^2 \del_T\del_{\bar T}\,-\,2\right]
    \Delta^{\rm univ}(T,U)\ &
=\ \left[ (U+\bar U)^2 \del_U\del_{\bar U}\,-\,2\right]
    \Delta^{\rm univ}(T,U)\cr
&=\ {1\over8\pi^2}\,\log\left| j(iT)\,-\,j(iU)\right|^2 .\cr
}
\eeq
{}From this equation, one may directly show that the $\Delta^{\rm univ}$
{\em has} to have the form (\ref{MIGSt}), where $\Ll(T,U)$ is precisely
as in eq.~(\ref{Lldef}), while $\foneol(T,U)$ is a holomorphic
function that transforms according to eqs.~(\ref{Tonh}) and has no
singularities except at $T\equiv U\!\!\pmod{SL(2,{\bf Z})}$ where
$\del_T\del_U\foneol$ has a logarithmic divergence (\ref{logDJ}).
This information is in turn sufficient to derive eqs.~(\ref{hfinal})
without any further field-theoretical input.
In this way, it is possible to obtain the Green-Schwarz term for
the toroidal compactifications of six-dimensional vacua of the
heterotic string (and, subsequently, of the $N=1$ orbifolds of such
compactifications) without using {\em any} special properties of the
$N=2$ supersymmetry but relying only on the string theory and
on the $N=1$ arguments of ref.~\cite{KLb}.

We conclude this article with a brief discussion of the Wilson-line
moduli $\phi^i$ which deform a toroidal compactification of a
six-dimensional vacuum and break some of its gauge symmetries.
For the sake of notational simplicity, we concentrate on a deformation
involving a single Wilson-line modulus which we denote as simply $\phi$;
the deformations involving several such moduli can be
analyzed in a similar manner.
The deformation reduces the gauge group $G$ of the un-deformed
theory to a subgroup $G'\subset S$; for small values of $\phi$
this reduction can be described in field-theoretical terms as a
Higgs mechanism in which $\phi$ plays the role of the Higgs field.
However, for $\left|\langle\phi\rangle\right|\not\ll1$,
one should simply integrate out the massive fields from the low-energy
EQFT; in the resulting ``deformed'' EQFT, $\phi$ becomes a
spectrum-preserving modulus.
As discussed in section~3.2, for moderately small values of $\phi$
the prepotential of this ``deformed''
EQFT is governed by the eqs.~(\ref{DFA}) and (\ref{FONEphi}),
which for the case at hand  give us
\beq
\eqalign{
\fone'(S,T,U,\phi)\ =\ {}&
S(\phi^2-TU)\ +\ \foneol(T,U)\ +\ \hath(T,U) \, \phi^2\cr
& -\ {b_{(\phi)}\over8\pi^2}
    \Big[ \log\phi+\log\eta^2(iT)+\log\eta^2(iU)
	+{\rm const}\Big]\phi^2\
    +\ \cdots\,,\cr
\noalign{\vskip 10pt}
\ftwo'_\ind(S,T,U,\phi)\ =\ {}&
S\ +\ \hath(T,U)\ +\ {b'_\ind\over8\pi^2}\,\log\phi\cr
& -\ {b_\ind\over8\pi^2}\Big[ \log\phi+\log\eta^2(iT)
	+\log\eta^2(iU)+{\rm const}\Big]\
    +\ \cdots\,, \cr
}\label{deformed}
\eeq
where the primes denote the parameters of the deformed theory,
$b_\ind$ and $b_{(\phi)}$ are the appropriate
beta-function coefficients
of the un-deformed theory for $\langle\phi\rangle=0$ and the
functions $\foneol(T,U)$ and $\hath(T,U)$ also belong to the
un-deformed theory ({\it cf.}~eqs.\ (\ref{hfinal}), (\ref{hathdef})
and (\ref{Lldef}) and also eqs.~(\ref{ffdet}) for the gauge couplings
of the un-deformed theory).

The `$\cdots$' in eqs.~(\ref{deformed}) stand for the sub-leading
terms carrying higher powers of the Wilson-line modulus $\phi$.
Such terms are severely constrained by the discrete modular
symmetries of the string theory.
In particular, several modular symmetries are common to all
Wilson-line deformations of toroidal compactifications, namely
the $T$-duality, the $U$-duality and the ``parities''
$T\leftrightarrow U$ and $\phi\to{-\phi}$.
The transformation rules for the $T$-duality
$SL(2,{\bf Z})_T$ of the deformed theory are
\beq
\eqalign{
T\ \to\ {aT-ib\over icT+d}\,,&
\qquad U\ \to\ U\ -\ {ic\phi^2\over icT+d}\,,\qquad
    \phi\ \to\ {\phi\over icT+d}\,,\cr
{\foneol}'(T,U,\phi)\ &
\to\ {{\foneol}'(T,U,\phi)+\shift(T,U,\phi)\over (icT+d)^2}\,,\cr
\ftwo'_\ind(S,T,U,\phi)\ &
\to \ftwo'_\ind(S,T,U,\phi)\
    -\ {b'_\ind\over8\pi^2}\,\log(icT+d)
+\ {\rm const.}\cr
}\label{DeformedRules}
\eeq
({\it cf.}\ eqs.\ (\ref{Tduality})
and (\ref{Hshift})--(\ref{fcomp}));
similar transformation rules with $U$ and $T$ interchanged
describe the $U$-duality $SL(2,{\bf Z})_U$ while
the parities $T\leftrightarrow U$ and $\phi\to{-\phi}$
leave the pre\-potential invariant.
Our task therefore is to solve for holomorphic functions
${\foneol}'(T,U,\phi)$ and $\ftwo'_\ind(S,T,U,\phi)$
which satisfy these transformation rules exactly and whose
small-$\phi$ limits are precisely as in eqs.~(\ref{deformed}).
For that purpose, let us first define the following functions:
\beq
\eqalign{
\Omega_4(T,U,\phi)\ &
= \sum_{n=0}^\infty {\phi^{2n+4}\over n!\,(n+3)!}
    \left(\del_T^n\,E_4(iT)\right) \left(\del_U^n\,E_4(iU)\right) ,\cr
\Omega_6(T,U,\phi)\ &
= \sum_{n=0}^\infty {\phi^{2n+6}\over n!\,(n+5)!}
    \left(\del_T^n\,E_6(iT)\right) \left(\del_U^n\,E_6(iU)\right) ,\cr
\Omega_{12}(T,U,\phi)\ &
= \sum_{n=0}^\infty {\phi^{2n+12}\over n!\,(n+11)!}
    \left(\del_T^n\,\eta^{24}iT)\right)
    \left(\del_U^n\,\eta^{24}(iU)\right) .\cr
}\label{ThreeOmegas}
\eeq
These three functions are $SO(2,2,{\bf Z})$ invariant,
holomorphic, non-singular throughout the $(T,U,\phi)$ moduli space
($({\rm Re}\, T)({\rm Re}\, U)>({\rm Re}\,\phi)^2$) and do not
grow faster than powers
of the moduli in any decompactification limit.
In other words, they are modular forms of the  $SO(2,2,{\bf Z})$
and furthermore, all such modular forms are polynomials or power
series in the $\Omega_4$, $\Omega_6$ and $\Omega_{12}$.
We also need three additional holomorphic functions:
\beq
\!\eqalign{
\Omega_A = \sum_{n=1}^\infty {\phi^{2n}\over (n!)^2} &
\,n(\del_T\del_U)^n\,\log\left( j(iT)-j(iU)\right),\cr
\Omega_L = \sum_{n=0}^\infty {\phi^{2n}\over (n!)^2} &
\left( (\del_T\del_U)^n\, +\,n(\del_T\del_U)^{n-1}
\left[\del_T\del_U ,\log(\eta^2(iT)\eta^2(iU))\right]\right)
\log\left( j(iT)-j(iU)\right) ,\cr
\Omega_H = \sum_{n=0}^\infty {\phi^{2n}\over (n!)^2} &
\left( \textstyle{1\over2}(n^2-6n+2) (\del_T\del_U)^n\right.\cr
&\quad\left.{}+\,n(n-1)(n-2)(\del_T\del_U)^{n-3}
\left[(\del_T\del_U)^3 ,\log(\eta^2(iT)\eta^2(iU))\right]\right)
\foneol(T,U) ,\cr
}\label{OmegaLH}
\eeq
where in the last definition $\foneol(T,U)$ is precisely as
defined by eqs.~(\ref{hfinal})
for the un-deformed toroidal compactifications.
These functions are singular along the critical $T\equiv U$ lines;
also, $\Omega_A(T,U,\phi)$ is modular invariant,
$\Omega_L(T,U,\phi)$ is modular invariant up to a constant imaginary
shift while
$\Omega_H(T,U,\phi)$ transforms precisely as the one-loop deformed
prepotential ${\foneol}'(T,U,\phi)$ should transform according to
(\ref{DeformedRules}).

With the above definitions (\ref{ThreeOmegas}) and (\ref{OmegaLH})
we can now write the general solution for
moduli prepotential and the gauge couplings of the deformed theory
that have both the right modular transformation properties
(\ref{DeformedRules}) and the right $\phi\to0$ limits (\ref{deformed}):
\beq
\eqalign{
{\foneol}'(T,U,\phi)\ &
=\ \Omega_H\
    -\ {\phi^2\over 8\pi^2}\,\left[\Omega_L\,
	-\,{(1-2\Omega_A)^2\over\Omega_A(1-\Omega_A)}\,
	\log{1-2\Omega_A\over1-\Omega_A}\right]\
    -\ {b_{(\phi)}\,\phi^2\over 96\pi^2}\,\log\Omega_{12}\cr
&\qquad+\sum_{k,\ell, m\ge0} C^h_{k\ell m}\,
	\Omega_4^k \Omega_6^\ell \Omega_{12}^m\,\phi^2 ,\cr
\ftwo'_\ind(S,T,U,\phi) &
=\ \hatS\ +\ {b'_\ind\over 8\pi^2}\,\log\phi\
    -\ {b_{(\phi)}\over 96\pi^2}\,\log\Omega_{12}\cr
&\qquad+\sum_{k,\ell, m\ge0} C^\ind_{k\ell m}\,
	\Omega_4^k \Omega_6^\ell \Omega_{12}^m\,,\cr
}\label{Wilsolutions}
\eeq
where
\beq
\eqalign{
\hatS(S,T,U,\phi)\ &
=\ S\ +\ {\textstyle {1\over10}}\left[ \Ll'(T,U,\phi)\,
	-\,4\del_T\del_U\,{\foneol}'(T,U,\phi)\,
	+\,\del_\phi^2\,{\foneol}'(T,U,\phi)\right] \cr
& =\ S\ +\ \hath(T,U)\ +\ O(\phi^2)\qquad \hbox{(for small $\phi$)}\cr
{\rm for}\quad\Ll(T,U,\phi)\ &
=\ {-1\over\pi^2}\left[\Omega_L(T,U,\phi)\,
	+\,\log{1-2\Omega_A\over1-\Omega_A}\right]
    +\ {b_{(\phi)}\over48\pi^2}\,\log\Omega_{12}(T,U,\phi)\cr
}
\eeq
and $C^h_{k\ell m}$ and $C^\ind_{k\ell m}$
are arbitrary complex constants.
Note however that in the small $\phi$ limit, $\Omega_4=O(\phi^4)$,
$\Omega_6=O(\phi^6)$ and $\Omega_{12}=O(\phi^{12})$.
Hence, the solution (\ref{Wilsolutions}) uniquely determines
the moduli-dependent gauge couplings $\ftwo'$ up to terms of
the order $\phi^4$ or higher and the moduli prepotential
$\fone$ up to the $O(\phi^6)$ or higher-order terms. Also note, that the
lowest terms in the expansions   (\ref{ThreeOmegas}), (\ref{OmegaLH})
and (\ref{Wilsolutions}) agree with previous results \cite{CLM1,ANTO}
on threshold corrections with Wilson lines.

The coefficients $C^h_{k\ell m}$ and $C^\ind_{k\ell m}$
cannot be determined by the $SO(2,2,{\bf Z})$ modular symmetries
that preserve the un-deformed subspace $\phi=0$ of the deformed
$(T,U,\phi)$ moduli space.
Instead, one should demand the correct transformation properties
of the prepotential under the entire symmetry group of the deformed
moduli space, namely $SO(2,3,{\bf Z})$.
In particular, all the gauge couplings should be periodic functions
of $\phi$.
Note however that the period of the Wilson-line modulus $\phi$
depends on the particular modulus.
To be precise, the orientation of the $(\P^0,\ldots,\P^4)$ basis
corresponding to the physical moduli $T$, $U$ and $\phi$ of the
deformed theory relative to the crystallographic basis of the discrete
$SO(2,3,{\bf Z})$ symmetry depends on a particular Wilson-line
deformation under consideration.
Consequently, at the $O(\phi^4)$ level,
the dependence of the gauge couplings on the
Wilson-line modulus $\phi$ depends on a particular modulus and
in models with several Wilson-line moduli $\phi^i$,
the $O(\phi^4)$ terms in the gauge couplings
generally have non-trivial index structure
($i.\,e.,\ f^\ind_{ijkl}(T,U)\phi^i\phi^j\phi^k\phi^l$).

In conclusion, the classical target-space duality symmetries
together with  informations about the singularity structure of
the gauge couplings allowed us to completely determine the
holomorphic prepotential for toroidal compactifications
to all orders in the perturbation theory.
(Up to the ambiguity
encoded in the $\shift(T,U)$ term, which amounts to a field-independent
ambiguity of some $\theta$ angles.)
This result leads to the Green-Schwarz mixing of the dilaton and the
moduli in the K\"ahler potential, which can be independently
confirmed by a direct string-loop computation.
Our analysis extends to moderately small Wilson-line deformations of
toroidal compactifications, for which we obtained the
model-independent leading terms in the expansion of the prepotential
into powers of the Wilson-line moduli.
We believe that our results are useful for the eventual
non-perturbative analysis of the $N=2$ vacua of the heterotic string
along the lines of refs. \cite{SeibWit,LKTY}.

\vskip 2cm

\noindent {\bf Acknowledgments: }

\noindent
We would like to thank D.~Anselmi, G.~Lopes Cardoso,
P.~Candelas, T.~Mohaupt, J.~Sonnenschein,
N.~Seiberg, S.~Theisen,  A.~Van Proeyen, C.~Vafa, E.~Witten
and S.~Yankielowicz for helpful discussions.

\noindent
B.~d.W.\ thanks the Humboldt-Universit\"at Berlin for hospitality and the
Andrejewski Foundation for financial support.
J.~L.\ thanks E.~Witten and the Institute for Advanced Study
as well as the Humboldt-Universit\"at  Berlin
for their hospitality.
V.~K.\ thanks the hospitality of the theory groups at
Universit\"at M\"unchen and Humboldt-Universit\"at  Berlin.

\noindent
The research of V.~K.\ is supported in part by the NSF,
under grant PHY--90--09850,
and by the Robert A.~Welch Foundation.
The research of J.~L.\ is supported by the Heisenberg Fellowship
of the DFG.
The collaboration of  V.~K.\ and J.~L.\ is additionally supported by
NATO, under grant CRG~931380. The collaboration of B.~d.W.\ and
J.~L.\ is part of the European Community Research Programme under contract
SC1--CT92--0789.

\setcounter{section}{0}
\setcounter{equation}{0}

\renewcommand{\thesection}{Appendix \Alph{section}.}
\renewcommand{\theequation}{A.\arabic{equation}}
\section{}
\vskip .8cm
\centerline{\bf
Dilaton-$B_{\mu\nu}$ $N=2$ Vector-Tensor supermultiplet}
\vskip .5cm

As mentioned in section 2, the $N=2$ heterotic string compactification
gives rise to a new supermultiplet consisting of a scalar $\phi$
(corresponding to the dilaton), a rank-two tensor gauge field
$B_{\mu\nu}$, a vector gauge field $V_\mu$ and a doublet of
Majorana spinors $\lambda_i$. These
fields describe an on-shell supermultiplet of two spin-0, one spin-1 and
four spin-$1/2$ states, which are also described by an $N=2$
vector multiplet. Therefore we
expect that the vector-tensor multiplet can be  converted into a
vector multiplet by means of a duality transformation.

The off-shell structure of the vector-tensor multiplet differs
from that of the vector multiplet in several respects. Off-shell
counting reveals that the $8+8$ field components can only be
realized as an off-shell supermultiplet in the presence of a
central charge. On shell this
central charge vanishes. In the context of {\em local}
supersymmetry the central charge must be gauged, and for that one
needs at least one Abelian vector multiplet (whose corresponding
gauge field could coincide with the graviphoton).
Although the central charge acts in a rather subtle way on the components
of the vector-tensor multiplet, we expect that its coupling to
$N=2$ supergravity can constructed along the same lines as
that for scalar (hyper)multiplets. Here we confine ourselves to
the linearized treatment of the multiplet.

The bosonic fields given above comprise only seven degrees of
freedom. The missing degree of freedom is provided by a real
scalar auxiliary field, which we denote by $D$. The linearized
transformation rules of the vector-tensor multiplet are as
follows,
\begin{eqnarray}
\delta\phi &=& \bar \epsilon^i \lambda_i +\bar \epsilon_i
\lambda^i \,,\nonumber\\
\delta V_\mu &=& i\varepsilon^{ij}\bar \epsilon_i\gamma_\mu
\lambda_j - i\varepsilon_{ij}\bar \epsilon^i\gamma_\mu\lambda^j\,
,\nonumber\\
\delta B_{\mu\nu} &=&2\bar \epsilon^i\sigma_{\mu\nu}\lambda_i
+2\bar \epsilon_i\sigma_{\mu\nu} \lambda^i\,,\\
\delta\lambda_i &=& (\partial\llap /\phi -iH\!\llap /\,)\epsilon_i
-\varepsilon_{ij}\,(i\sigma\cdot F^- +D)\epsilon^j\,,\nonumber\\
\delta D &=&  \varepsilon^{ij}\,\bar \epsilon_i\partial\llap /
\lambda_j  + \varepsilon_{ij}\,\bar \epsilon^i\partial\llap /
\lambda^j\, . \nonumber
\end{eqnarray}
We use the chiral notation employed in \cite{DWVHVP,DWLVP}, where, for
spinor quantities, upper and
lower $SU(2)$ indices $i,j,\ldots$ denote chiral components.
For the spinors used above the precise correspondence is
$$
{\eqalign{\gamma_5 \lambda_i &= \lambda_i \,,\cr
\gamma_5 \lambda^i &= -\lambda^i\,, \cr}}
\qquad{\eqalign{\gamma_5 \epsilon^i &= \epsilon^i \,,\cr
\gamma_5 \epsilon^i &= - \epsilon^i\,. \cr}}
$$
The $SU(2)$ indices are raised and lowered by complex
conjugation.
The quantities $H^\mu$ and $F^\pm_{\mu\nu}$ are the field strength of the
tensor field and the (anti)selfdual field strengths of the vector
field, defined by
\begin{eqnarray}
H^\mu &=& {\textstyle{1\over 2}} i
\varepsilon^{\mu\nu\rho\sigma}\partial_\nu B_{\rho\sigma}\,,
\nonumber\\
F_{\mu\nu} &=& F_{\mu\nu}^+ + F_{\mu\nu}^- = \partial_\mu
V_\nu-\partial_\nu V_\mu\,.
\end{eqnarray}
They satisfy the Bianchi identities
\begin{equation}
\partial_\mu H^\mu = 0\,,\qquad \partial^\mu F_{\mu\nu}^+
=\partial^\mu F_{\mu\nu}^-\,.
\end{equation}
For completeness we also record their supersymmetry transformations
\ber
\delta F^-_{\m\n} &=& i\varepsilon^{ij}\bar
\epsilon_i\partial\llap/ \s_{\mu\n} \lambda_j +
i\varepsilon_{ij}\bar \epsilon^i\s_{\m\n}\partial\llap/\lambda^j\,
,\nonumber\\
\delta H^\mu &=&-2i\bar
\epsilon^i\sigma^{\mu\nu}\partial_\n\lambda_i +2i \bar
\epsilon_i\sigma^{\mu\nu} \partial_\n\lambda^i\,.
\eer{}

The supersymmetry algebra closes on the above fields. The
anticommutator of two supersymmetry transformations leads to
a general coordinate transformation, central charge transformations
and gauge transformations on the vector and tensor gauge fields.
The central charge transformations take the form
\begin{eqnarray}
\delta_{\rm z}\phi &=& -{\textstyle{1\over 2}} (z+\bar z)\, D \,,
\nonumber\\
\delta_{\rm z} V_\mu &=& {\textstyle{1\over 2}}(z+\bar z)\,H_\mu
\, ,\nonumber\\
\delta_{\rm z} B_{\mu\nu} &=&i z\,F_{\mu\nu}^- - i\bar z\,
F_{\mu\nu}^+ \,,\label{ztrans}\\
\delta_{\rm z}\lambda_i &=& -{\textstyle{1\over 2}}(z+\bar z)\,
\varepsilon_{ij}\,\partial\llap/\lambda^j \,,\nonumber\\
\delta_{\rm z} D &=&  {\textstyle{1\over 2}}(z+\bar z) \,
\partial^2\phi\,.  \nonumber
\end{eqnarray}
In the supersymmetry commutator $[\delta(\epsilon_1),
\delta(\epsilon_2)]$ the
central-charge parameter $z$ equals $z =4\bar
\epsilon_2^i\epsilon_1^j\,\varepsilon_{ij}$; the general-coordinate
transformation is given by $\xi^\m=
2(\bar\e_2^i\g^\m\e_{1i} + \bar\e_{2i}\g^\m\e_1^i)$.

{}From the product of two  vector-tensor multiplets,
one constructs an $N=2$ linear multiplet with central charge. In
components this linear multiplet is given by
\begin{eqnarray}
L_{ij}&=& \bar\lambda_i \lambda_j +
\varepsilon_{ik}\varepsilon_{jl}\,\bar\lambda^k\lambda^l \,,\nonumber\\
\varphi^i&=& -(\partial\llap/\phi +iH\!\llap /\,)\lambda^i +
\varepsilon^{ij} (-i\sigma\cdot F^-+D)\lambda_j \,,\nonumber\\
G&=& (\partial\phi)^2 - H^2 -2i H\cdot\partial\phi + (F^-)^2
-D^2  +\bar\l^i\partial\llap/ \l_i +\bar\l_i\partial\llap/ \l^i
 \,,\\
E_\mu&=&2H^\n (F^++F^-)_{\n\m}-2i \partial^\n\!\phi\, (F^+-F^-)_{\n\m}+2D\,
\partial_\m\phi \nonumber \\
&& + \varepsilon^{ij} \l_i\partial_\m \l_j  + \varepsilon_{ij}
\l^i\partial_\m \l^j  \,,\nonumber
\end{eqnarray}
{}From (\ref{ztrans}) it is straightforward to obtain the central
charge transformations of the linear multiplet components. For
example,
\begin{eqnarray}
\delta_{\rm z}L_{ij} &=&  (z+\bar z)\,
\varepsilon_{k(i}(\bar\l_{j)}\partial\llap/
\l^k - \bar\l^k\partial\llap/ \l_{j)}  ) \,,
\nonumber\\
\delta_{\rm z} G &=& (z+\bar z)\,\partial^\m\big[ -D(\partial_\m
\phi +iH_\m) + 2H^\n  F^-_{\n\m} -2i\phi\,\partial^\n \!F^-_{\n\m}
\nonumber \\
&&\hspace{20mm}  - {\textstyle{1\over 2}}
\varepsilon^{ij}\bar \l_i\partial_\m\l_j  -{\textstyle{1\over
2}}\varepsilon_{ij} \bar\l^i\partial_\m \l^j \big] \, .\label{ztransl}
\end{eqnarray}
The second equation shows that the appropriate constraint for
the linear multiplet is satisfied \cite{DWVHVP},
\begin{equation}
(z+\bar z) \,\partial_\m E^\m = -\delta_{\rm z} (G +\bar G)\,.
\end{equation}

The real part of $G$ yields a linearized supersymmetric
Lagrangian for the vector-tensor multiplet
\begin{equation}
{\cal L} = -{\textstyle{1\over 2}}(\partial\phi)^2
-\bar \lambda^i\partial\llap/ \lambda_i +
{\textstyle{1\over 2}}H^2 - {\textstyle{1\over 4}}F^2
+{\textstyle{1\over 2}}D^2   \,.    \label{vtaction}
\end{equation}
The other components of the linear multiplet play a role when
considering the invariant action in the background of a vector
multiplet that gauges the central charge.

As far as its physical degrees of freedom are concerned,  the
action (\ref{vtaction}) describes the same states as the action
for a vector
multiplet. In components this is rather obvious, as one can, by
means of a duality transformation,
convert the antisymmetric tensor field $B_{\m\n}$ into a
(pseudo)scalar field. The latter can be combined with the field
$\phi$ into a complex scalar field. At present it is not clear
how to perform the duality transformation in a way that is
manifestly supersymmetric off shell. The fact that only one of
the two multiplets has a central charge would certainly be a
nontrivial aspect of such a duality transformation.
It is worth mentioning that there exists also an $N=2$
tensor multiplet, consisting of three scalars, an antisymmetric
tensor gauge field, a doublet of spinors and a complex auxiliary
field, which can be converted to a scalar (hyper)multiplet by a
duality transformation \cite{DWPVP}. Again, one of the two
multiplets involved in the duality transformation has an off-shell
central charge. Also in
this case it is not yet known how to perform the duality
transformation such that supersymmetry is manifest off shell.
As discussed in the previous section, the $N=2$ tensor multiplet
arises in Calabi-Yau compactifications of
type-II superstrings.

In principle, by studying the supergravity and Chern-Simons
couplings of the new multiplet, one should be able to elucidate
the restrictions imposed on the dilaton-$B_{\m\n}$ system as
described in the dual formulation in terms of a vector multiplet.
This is an interesting topic, which deserves further study.
The strategy of this paper is to work in the dual formulation
and use all possible information from string theory to
specify the couplings of the corresponding vector multiplet.
Therefore the thrust of our work is on vector multiplets.

\end{document}